\documentclass[a4paper,11pt]{article}
\usepackage[margin=2.5cm]{geometry}
\usepackage{setspace,graphicx,url,hyperref,enumitem}
\usepackage[linesnumbered,ruled]{algorithm2e}
\usepackage{amsmath,amssymb,amsfonts,amsthm,natbib}

\begin{document}
\onehalfspacing
\title{Decomposition algorithms for solving NP-hard problems on a quantum annealer}
\author{Elijah Pelofske, Georg Hahn, and Hristo Djidjev}
\date{Los Alamos National Laboratory}
\maketitle

\begin{abstract}
NP-hard problems such as the maximum clique or minimum vertex cover problems, two of Karp's 21 NP-hard problems, have several applications in computational chemistry, biochemistry and computer network security. Adiabatic quantum annealers can search for the optimum value of such NP-hard optimization problems, given the problem can be embedded on their hardware. However, this is often not possible due to certain limitations of the hardware connectivity structure of the annealer. This paper studies a general framework for a decomposition algorithm for NP-hard graph problems aiming to identify an optimal set of vertices. Our generic algorithm allows us to recursively divide an instance until the generated subproblems can be embedded on the quantum annealer hardware and subsequently solved. The framework is applied to the maximum clique and minimum vertex cover problems, and we propose several pruning and reduction techniques to speed up the recursive decomposition. The performance of both algorithms is assessed in a detailed simulation study.
\end{abstract}

\section{Introduction}
\label{sec:intro}
Novel computing technologies allows one to search for solutions of NP-hard (graph) problems that are very hard to solve classically~\citep{Chapuis2017}. One such device is the quantum annealer of D-Wave Systems, Inc.~\citep{TechnicalDescriptionDwave}, which can propose approximate solutions of quadratic unconstrained binary optimization (QUBO) and Ising problems given by the minimum of a function of the form
\begin{align}
    H(x_1,\ldots,x_n) = \sum_{i=1}^n a_i x_i + \sum_{i<j} a_{ij} x_i x_j.
    \label{eq:ising}
\end{align}
In \eqref{eq:ising}, the coefficients $a_i \in \mathbb{R}$, $i \in \{1,\ldots,n\}$, are linear weights and $a_{ij} \in \mathbb{R}$ for $i<j$ are quadratic weights. The problem \eqref{eq:ising} is called a QUBO problem if $x_i \in \{0,1\}$ and an Ising problem if $x_i \in \{-1,+1\}$ for all $i$. The function \eqref{eq:ising} is often called a QUBO or Ising function, respectively. The formulation in \eqref{eq:ising} is general enough to allow all NP-hard problems to be formulated as minimizations of such a function \citep{Barahona1982}. Both QUBO and Ising formulations are equivalent~\citep{Barahona1982, Choi2008, Lucas2014, Djidjev2016EfficientAnnealing}. The D-Wave quantum annealer aims to find a minimum of the function \eqref{eq:ising} by mapping it to a physical quantum system, from which a solution is read off after hardware-implemented annealing is completed. In such a mapping, linear weights are mapped onto qubits and quadratic weights are mapped onto links between qubits called \textit{couplers}. Moreover, if $a_i$ and $a_j$ are mapped onto qubits $q_i$ and $q_j$, then $a_{ij}$ is mapped onto the coupler connecting $q_i$ and $q_j$.

However, directly computing a minimum of a given function of type \eqref{eq:ising}  on a quantum annealer is often not possible due to a variety of reasons: first, there is a limitation on the input problem size that can fit on the quantum hardware due to the finite number of available qubits (up to roughly 2000 qubits for the newest D-Wave 2000Q\texttrademark~computer). Second, even if the number of qubits exceeds the number of variables, the current (D-Wave) technology provides only limited qubit connectivity \citep{TechnicalDescriptionDwave}. It is thus not guaranteed that all the required quadratic couplers needed to map a specific problem onto the annealer hardware are available. This problem can be alleviated with a so-called \textit{minor embedding} of the problem function onto the D-Wave hardware, where each variable is mapped onto a set of connected qubits, rather than a single one, at the expense, however, of a severe reduction in the number of available qubits~\citep{Choi2008, Chapuis2017}. For instance, the largest embeddable QUBO (of arbitrary connectivity) on D-Wave 2000Q has $64$ variables, thus guaranteeing that arbitrary QUBO problems with up to $64$ variables can be approximately solved on D-Wave. For QUBOs with a sparse connectivity structure, tailored embeddings can allow for larger instances to be solved on D-Wave. We note that quantum annealers such as D-Wave do not provide a guarantee of correctness, and thus typically return approximate solutions (of high quality).

In this article we propose a general decomposition algorithm for NP-hard graph problems aiming to find an optimal set of vertices by minimizing \eqref{eq:ising}. The proposed approach makes it possible to solve problems on D-Wave with sizes exceeding the number of available qubits. The decomposition algorithm recursively splits a given instance into smaller subproblems until, at a certain recursion level, the generated subproblems are small enough to be solved directly, e.g., using a quantum annealer. The decomposition algorithm is exact, meaning that the optimality of the solution is guaranteed provided all generated subproblems are solved exactly.

We apply our decomposition technique to two important NP-hard problems: the maximum clique (MC) and the minimum vertex cover (MVC) problems. Formally, we are given an undirected graph $G=(V,E)$ with vertex set $V$ and edge set $E \subseteq V \times V$. A subgraph $G(W)$ of $G$ induced by a subset $W \subseteq V$ is called a \textit{clique} if there exists an edge in $E$ between any two vertices in $W$, and $G(W)$ is  a \textit{maximum clique} if $G(W)$ is a clique of maximum size. A subset $U \subseteq V$ is called a \textit{vertex cover} if every edge in $E$ has at least one endpoint in $U$, that is, for every $e=(u,v) \in E$ it holds true that $u \in U$ or $v \in U$. A \textit{minimum vertex cover} is a vertex cover of minimum size.

It is known that all NP-hard problems can be expressed as the minimization of a function of the form \eqref{eq:ising} including, for instance, the graph partitioning, the graph coloring, or the maximum clique problems: see \cite{Lucas2014} for a comprehensive overview of QUBO and Ising formulations for a variety of NP-hard problems. For instance, the QUBO formulation for solving MC on a graph $G=(V,E)$ is given by
\begin{align}
    H_{MC} = -A\sum_{v \in V} x_v + B\sum_{(u,v) \in \overline{E}} x_u x_v,
    \label{eq:MC}
\end{align}
where the constants can be chosen as $A=1$, $B=2$ \citep{Lucas2014}. Analogously, for solving MVC on a graph $G=(V,E)$, we consider
\begin{align}
    H_{MVC} = A' \sum_{(u,v) \in E} (1-x_u)(1-x_v) + B' \sum_{v \in V} x_v,
    \label{eq:MVC}
\end{align}
where $0<B'<A'$ is required in order to ensure that minimizing \eqref{eq:MVC} is equivalent to solving the MVC problem. As an explicit choice, we fix $B'=1$ and $A'=2$ in the remainder of the article. In both \eqref{eq:MC} and \eqref{eq:MVC}, each $x_v \in \{0,1\}$ for $v \in V$ is a binary variable indicating if vertex $v$ belongs to the MC or the MVC, respectively.

Though proven to be exact, the decomposition algorithms we present in this work have a worst-case exponential runtime (which is to be expected since the problems we are solving are NP-hard). We therefore aim to reduce the amount of computations as much as possible. To this end, a variety of techniques outlined in Section~\ref{sec:pruning} allows one to eliminate a large number of subproblems during the recursion that cannot contribute to MC or MVC, or to reduce the size of some subproblems by removing vertices which cannot belong to the optimal solution.

This article is a journal version of \cite{acmpaper}, published in the \textit{Proceedings of the 16th ACM International Conference on Computing Frontiers 2019}. In contrast to the conference paper, this journal version contains a much more general decomposition framework which is applicable to a broader class of NP-hard graph problems. We show how the proposed framework can be concretized into previously published decomposition algorithms for the MC problem \citep{Pelofske2019} and the MVC problem \citep{acmpaper} as special cases. We evaluate both methods simultaneously in a unified simulation section. Since the two problems are related to each other, we will be able to empirically highlight asymmetries between them in the analysis section.

This article is structured as follows. After a brief literature review in Section~\ref{sec:literaturereview}, Section~\ref{sec:algorithm} introduces our general decomposition framework, whose implementation we demonstrate for the MC (Section~\ref{sec:MC}) and MVC (Section~\ref{sec:MVC}) problems. To prune subproblems during the decomposition, we discuss a variety of bounds and reduction techniques in Section~\ref{sec:pruning}. We assess the performance of our decomposition methods in a detailed simulation study in Section~\ref{sec:experiments}. The article concludes with a discussion of our results in Section~\ref{sec:discussion}.

\subsection{Literature review}
\label{sec:literaturereview}
The development of exact algorithms for NP-hard problems has been an area of constant attention in the literature~\citep{dimacs1996, dimacs2000, Woeginger2008}.

In particular, the minimum vertex cover problem has been widely studied in the literature from a variety of aspects~\citep{DowneyFellows1992, Balasubramanian1998, StegeFellows1999, Chen2000, Chen2001, NiedermeierRossmanith2003}. For instance, \cite{Chen2010} present a selection of techniques to reduce the size of an MVC instance and introduce a polynomial space and exponential time algorithm of the order of $O(1.27^k)$, where $k$ is the sought maximal size of the MVC, thus improving over the $O(1.29^k)$ algorithm of \cite{Niedermeier20070UB}. A variation of the MVC problem, the weighted MVC problem, is studied in \cite{Xu2016}.

Decomposition algorithms, such as the algorithm presented in this article, have already been suggested in \cite{Tarjan1985} and successfully applied to solve a variety of NP-hard problems such as graph coloring, see \cite{Rao2008}.

For quantum annealing, a decomposition algorithm for the maximum clique problem has been proposed in \cite{Chapuis2017} and \cite{Pelofske2019}. In \cite{Pelofske2019}, the authors additionally investigate a variety of techniques to prune subproblems during the recursive decomposition, for instance by computing bounds on the clique size. Similarly, to solve the maximum independent set problem, an equivalent formulation of the maximum clique problem, several algorithms are known including some relying on graph decomposition \citep{Giakoumakis1997,Courcelle2000}.

The algorithm of \cite{BronKerbosch1973} solves a related problem, that is the problem of enumerating all maximum cliques in a graph. Parallel version of this algorithm are available in the literature \citep{Rossi2015}. The algorithm of \cite{CarraghanPardalos1990} is another exact method to partially enumerate all maximum cliques.

Further exact algorithms have been developed in recent year, see for instance \cite{Robson1986, Robson2001} and \cite{Xiao2013}, including those based on principles such as \textit{measure and conquer} \citep{Fomin2006}.

Another area of research are branch-and-prune heuristics, including algorithms which use different solvers when the subproblems are sufficiently small (as we do in this work), see \cite{Hou2014} and \cite{Morrison2016} for a survey.

\section{Decomposing NP-hard graph problems}
\label{sec:algorithm}
This section describes a generic algorithm to decompose an NP-hard optimization graph problem that aims to find an optimal set of vertices minimizing or maximizing a given objective function. Our basic algorithm targets problems with binary decisions for each vertex: for instance, in \eqref{eq:MC} and \eqref{eq:MVC}, we are interested in the value of the binary indicator $x_v$ for each $v \in V$ which encodes with $x_v=1$ that vertex $v$ belongs to the maximum clique (or the minimum vertex cover).

The aim of our decomposition is to split up a problem instance into two subproblems with the property that (a) both subproblems are strictly smaller than the original instance, and (b) solving each exactly allows to reconstruct the optimal solution of the original instance in polynomial time. Applying the decomposition recursively thus allows one to decompose a given problem instance into arbitrarily small subproblems.

\subsection{Generic algorithm}
\label{sec:generic_algo}
Algorithm~\ref{algo:decomposition} illustrates the general structure of our decomposition algorithms assuming the problem is of minimization type. We start with an input graph $G=(V,E)$, a current solution $S$ (initialized as the empty set), and a value $\mu$ of the objective function for the current solution. For instance, for MC, the set $S$ will be the set of vertices belonging to the maximum clique, and $\mu$ will be the clique size. If the graph problem under investigation is a minimization problem, we start with $\mu=\infty$. Additionally, we require some cutoff size $s_{\max}$, which determines the size at which we stop the decomposition as the subproblems are small enough to be solved with a quantum or classical method.

First, a vertex $v$ for splitting the solution space is selected. Possible choices investigated in this work are given in Section~\ref{sec:vertexchoice}. The splitting vertex $v$ is used to split $G$ into two graphs $G^+$ and $G^-$ on which the optimization problem (e.g., MC or MVC) will be solved, where the precise splitting routine is problem dependent. For graph $G^+$, we assume that $v$ belongs to the optimal solution (e.g., the maximum clique), and for $G^-$ we assume it does not. Sections~\ref{sec:MC} and \ref{sec:MVC} give specific implementations of the splitting techniques for the maximum clique and the minimum vertex cover problems, respectively.

Before decomposing $G^+$ and $G^-$ further in a similar fashion, we aim to reduce the computational burden in Algorithm~\ref{algo:decomposition} by using context-specific knowledge about the graph problem to compute lower and upper bounds on the solutions in $G^+$ and $G^-$ (see Section~\ref{sec:bounds} for a list of bounds we employ). If it is impossible that $G^+$ or $G^-$ contain a solution that improves upon $\mu$, the value of the best solution found so far, we can discard them. Otherwise, we again use context-specific knowledge to reduce the size of existing subproblems through vertex and edge removal techniques (see Section~\ref{sec:mc_reductions} for reduction techniques for MC and Section~\ref{sec:mvc_reductions} for MVC).

If any of the subgraphs ($G^+$ or $G^-$) contains more vertices than the cutoff value $s_{\max}$, the splitting is recursively called on that subgraph using the current values of $S$ and $\mu$. Otherwise, i.e., if the corresponding subgraph ($G^+$ or $G^-$) is within the size limit $s_{\max}$, we solve the optimization problem using any classic or quantum algorithm of choice and update the values of $S$ and $\mu$ appropriately. At the end of the recursion, $S$ and $\mu$ contain the correct values of the solution and the corresponding value of the objective function.

If a maximization problem ought to be solved, we initialize $\mu=-\infty$. In line~5 of Algorithm~\ref{algo:decomposition}, we discard problems $G^+$ or $G^-$ if improvement over $\mu$ is impossible, in the sense that the solution of the subproblems $G^+$ or $G^-$ will be less than $\mu$. In line~11, we return $S^+,\tilde{\mu}^+$ if $\tilde{\mu}^+ \geq \tilde{\mu}^-$.

\begin{algorithm}[t]
    \caption{\texttt{decomposition for a minimization problem}}
    \label{algo:decomposition}
    \SetKwInOut{Input}{input}
    \SetKwFor{Loop}{repeat}{}{end}
    \Input{$G=(V,E)$, $S \leftarrow \emptyset$, $\mu \leftarrow \infty$, $s_{\min}$, additional parameters\;}
    Choose $v \in V$ according to some selection criterion\;
    Denote by $S^*$ any optimal solution for $G$. Using context-specific knowledge about the NP-hard graph problem, define the following proper subgraphs of $G$:\\
    a) $G^+ = (V^+,E^+)$ such that $S^*$ is an optimal solution for $G^+$ if $v\in S^*$; Update $\mu^+$\;
    b) $G^- = (V^-,E^-)$  such that $S^*$ is an optimal solution for $G^-$ if $v\not\in S^*$; Update $\mu^-$\;
    c) Bound the value of the optimal solutions in $G^+$ and $G^-$, and discard any of them (set $G^+$ or $G^-$ to $\emptyset$) if improvement over $\mu$ is impossible\;
    d) Attempt to reduce the size of the subgraphs $G^+$ and $G^-$ through vertex and edge removal techniques\;
    \lIf{$|V^+| > s_{\min}$} {$S^+,\tilde{\mu}^+ =$ \texttt{decomposition}$(G^+,S \cup \{v\},\mu^+,s_{\min})$}
    \lElse{Solve graph problem directly on $G^+$ and update set of solution vertices $S^+$ and value $\tilde{\mu}^+$}
    \lIf{$|V^-| > s_{\min}$}{$S^-,\tilde{\mu}^- =$ \texttt{decomposition}$(G^-,S,\mu^-,s_{\min})$}
    \lElse{Solve graph problem directly on $G^-$ and update set of solution vertices $S^-$ and value $\tilde{\mu}^-$}
    \lIf{$\tilde{\mu}^+ \leq \tilde{\mu}^-$}{\Return~$S^+,\tilde{\mu}^+$}
    \lElse{\Return~$S^-,\tilde{\mu}^-$}
\end{algorithm}

Algorithm~\ref{algo:decomposition} does not require that subproblems are solved on a quantum device such as the D-Wave annealer. In principle, any suitable device or method can be used to exactly solve any of the generated subproblems at any stage of the decomposition. However, one straightforward choice is to stop decomposing a subproblem further once it can be embedded on the D-Wave hardware, that is once the subgraph size is at most $s_{\max}=46$ vertices for the D-Wave 2X at Los Alamos National Laboratory (the largest size of an arbitrary problem that can be embedded on the hardware). For the D-Wave 2000Q at Los Alamos National Laboratory, this cutoff is $s_{\max}=64$ vertices, and for D-Wave Advantage it is $s_{\max}=180$ vertices.

Algorithm~\ref{algo:decomposition} can also be applied probabilistically: If the solver applied to the subgraphs at leaf level finds optimal solutions with some probability $p$, our decomposition algorithm will report the correct solution for the original graph $G$ with a probability that is a function of $p$.

\subsection{Vertex Choice}
\label{sec:vertexchoice}
One tuning parameter of Algorithm~\ref{algo:decomposition} is the procedure for selecting the vertex $v$ that is used in each iteration to split the current graph instance $G$ into two new graphs. Possible choices include:
\begin{enumerate}
    \item a vertex $v$ of lowest degree.
    \item a vertex $v$ of median degree.
    \item a vertex $v$ of highest degree.
    \item a vertex $v$ chosen at random.
\end{enumerate}
In any of the above cases, if multiple vertices satisfy the selection criterion, the vertex $v$ which is extracted is chosen at random. The aforementioned vertex selection approaches are experimentally explored in Section~\ref{sec:experiments_vertexchoice}.

\subsection{Implementation for Maximum Clique}
\label{sec:MC}
\begin{figure}[t]
    \centering
    \includegraphics[width=0.5\textwidth]{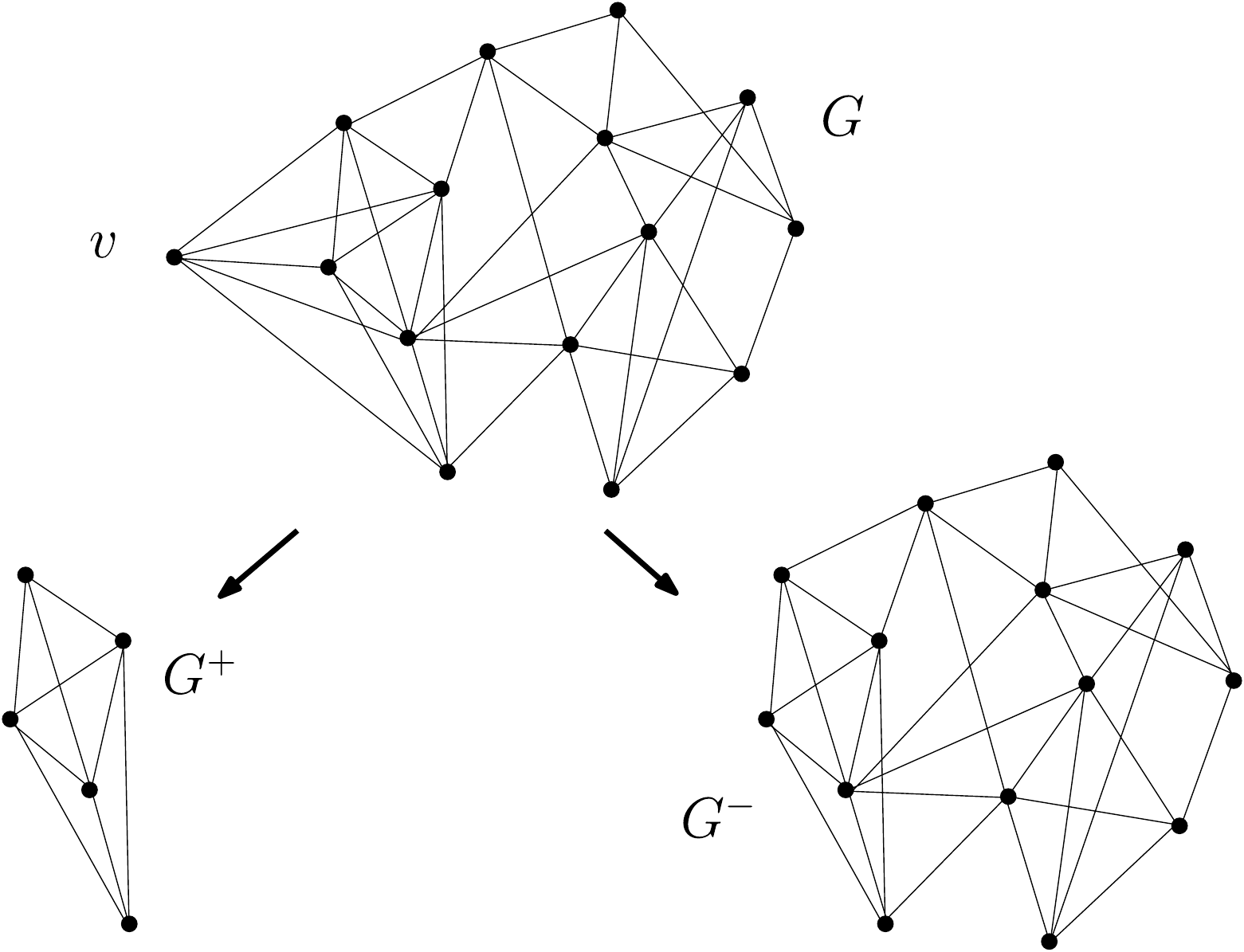}
    \caption{Illustration of the vertex splitting at a vertex $v$ for MC.\label{fig:mc_vertex_splitting}}
\end{figure}
Using Algorithm~\ref{algo:decomposition} as a framework, we now fill out the details of the generic implementation to arrive at a decomposition algorithm for MC.

We use the CH-partitioning introduced in \cite{Djidjev2015}, see also \cite{Chapuis2017}, in order to split a large input graph $G=(V,E)$ into smaller subgraphs on which a maximum clique is found. Denoting the unknown maximum clique as $V' \subseteq V$, there are two cases. Either $v \in V'$ or $v \notin V'$, each case leading to a subproblem of reduced size. If $v \in V'$, we extract the subgraph $G^+$ containing all neighbors of $v$ and all edges between them. We also set $\mu^+:=\mu+1$. If $v \notin V'$, the vertex $v$ and all edges adjacent to $v$ are removed from $G$, thus creating the graph $G^-$ (and $\mu^-:=\mu$). This is visualized in Figure~\ref{fig:mc_vertex_splitting}.

Since the cases $v \in V'$ and $v \notin V'$ are exhaustive, we continue to compute MC on both subgraphs $G^+$ and $G^-$. Since neither of the generated subgraphs $G^+$ and $G^-$ contains $v$, the graph size is reduced by at least one in each recursion level, thus guaranteeing the termination of the algorithm. The clique number $\omega(G)$ of $G$ is equal to $\min(\omega(G^+)+1,\omega(G^-))$.

\subsection{Implementation for Minimum Vertex Cover}
\label{sec:MVC}
\begin{figure}
    \centering
    \includegraphics[width=0.5\textwidth]{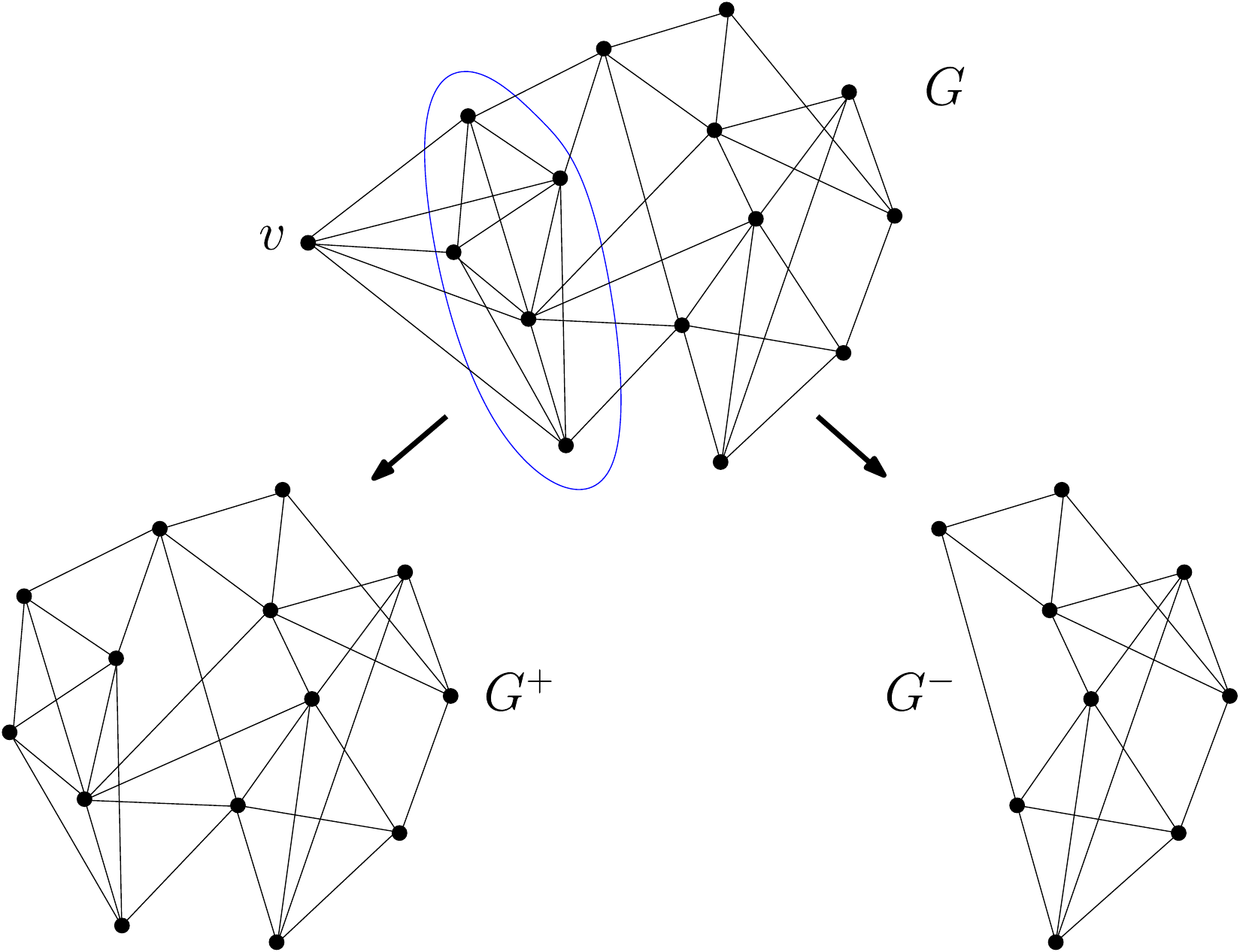}
    \caption{Illustration of the vertex splitting at a vertex $v$ for MVC.}
    \label{fig:mvc_vertex_splitting}
\end{figure}
We again select an arbitrary vertex $v \in V$ to split the input graph $G$, see Figure~\ref{fig:mvc_vertex_splitting}. Denoting the unknown MVC as $V' \subseteq V$, there are again two cases leading to subproblems of reduced sizes. If $v \in V'$, we add $v$ to the MVC, update $\mu^+$, and remove $v$ and all edges adjacent to $v$ from the graph since those edges are already covered by the choice of $v$. The resulting graph is $G^+$ as illustrated in Figure~\ref{fig:mvc_vertex_splitting}.

If $v \notin V'$, we observe that for all edges with endpoint $v$, that is for all $(v,u) \in E$, it must hold true that $u \in V'$ ($\mu^-$ is updated accordingly). This is true since, if $v \notin V'$, those edges must still be covered by their other endpoint $u$ in the MVC. Also, we can remove $v$ from $G$ since we know it is not in the MVC. Likewise, we can remove all $u$ with $(u,v) \in V$ and the adjacent edges of all such $u$ since those vertices are known to belong to the MVC. In Figure~\ref{fig:mvc_vertex_splitting}, under the assumption that $v \not\in V'$, all vertices inside the blue circle must belong to the MVC. After removing $v$ and all its adjacent edges, and assigning all encircled vertices to the MVC, we are left with the subgraph $G^-$.

As for the MC problem, the cases $v \in V'$ and $v \notin V'$ are exhaustive. In a recursive application, some bookkeeping is needed to keep track of the current set of cover vertices for each generated subgraph.

\subsection{Decomposition of other NP-hard problems}
\label{sec:other_NP_problems}
Apart from the MC and MVC problems considered in Sections~\ref{sec:MC} and \ref{sec:MVC}, the decomposition algorithm of Section~\ref{sec:generic_algo} can be applied to a much broader class of NP-hard problems. This section gives a brief overview of how such decompositions might work. We assume that any graph problem listed below is defined on some input graph $G=(V,E)$.
\begin{enumerate}
    \item Graph partitioning into two components is a classic NP-hard problem. One can decompose it by assuming that an edge is either a cut edge or it is not, implying that the two adjacent vertices are in different partitions or the same one.
    \item Graph coloring with $n$ colors has a QUBO formulation with $|V|\cdot n$ variables in which variable $x_{v,i}$ indicates with a $1$ if vertex $v$ has color $i \in \{1,\ldots,n\}$ \citep{Lucas2014}. We can decompose this problem easily by assigning a set of feasible colors to each vertex. We then select a random vertex $v \in V$ and probe each possible color $i$ for $v$, implying that color $i$ is removed from the available colors of all vertices adjacent to $v$.
    \item Hamiltonian Cycles can likewise be encoded with $|V|^2$ binary variables in which variable $x_{v,i}$ indicates with a $1$ the $i$th place of vertex $v \in V$ in the sorted order of visited vertices. We can decompose this problem into $|V|$ subproblems by selecting a random vertex and assigning it each possible rank in the sorted order of visited vertices. For each assigned rank, the possible ranks for all other vertices decrease by one, and moreover we know that one of the adjacent vertices to $v$ will have rank $i-1$, and another rank $i+1$.
    \item The Traveling Salesman problem is defined on a graph with edge weights. The optimal Traveling Salesman solution is a Hamiltonian Cycle of minimal total edge weight. Using the same indicators $x_{v,i}$ for $v \in V$, we can use the same strategy as the one for Hamiltonian Cycles to decompose such a problem.
\end{enumerate}

In general, any arbitrary QUBO or Ising model can be trivially decomposed using a technique called \textit{probing} in \cite{Boros2002}. For this, we select a random binary variable $x_i$ and create two new QUBOs (or Ising models) by setting $x_i:=0$ and $x_i:=1$ (or $-1$ and $+1$ in the case of Ising models). This will not only eliminate the quadratic term $x_i^2$ but also reduce all quadratic terms involving $x_i$ to linear terms. The resulting QUBO (or Ising model) can then be analysed using general purpose bounding or reduction techniques such as the ones in \cite{Boros2002} in order to prune subproblems.

\section{Pruning techniques for MC and MVC}
\label{sec:pruning}
The recursive decomposition proposed in Section~\ref{sec:generic_algo} allows us to specify problem-specific techniques to bound the optimal solution contained in the generated subproblems, and to reduce the size of the generated subproblems with the help of reduction techniques. The specific bounds and reductions we consider in the simulations are discussed in this section.

\subsection{Upper and lower bounds for MC and MVC}
\label{sec:bounds}
We bound the size of the MC and MVC of each generated subgraph. The MC and MVC problems are in a way complementary problems, as the sum of the size of the MVC of a graph $G$ and the size of the MC of the complement of $G$ equals the number of the vertices of $G$ (due to the fact that $C$ is a clique in $G$ if and only if $V\setminus C$ is a vertex cover in the complement of $G$). Hence, a lower bound technique on one can be used for an upper bound on the other, and vice versa. As an example for using bounds, in the case of MVC, if a lower bound on the vertex cover in any subgraph is greater than or equal to the current best vertex cover size, we do not need to consider that subproblem for further decomposition as it cannot contain a better solution than the one already known.

The following are \textit{upper bounds} for MC and \textit{lower bounds} for MVC:
\begin{enumerate}
    \item First, we take the minimum (for MC) or maximum (for MVC) of three easy to compute bounds.
        \begin{enumerate}
            \item The function \textit{min\_weighted\_vertex\_cover} of the NetworkX package \citep{Networkx} computes an approximate vertex cover of at most twice the size of the optimal cover using the algorithm of \cite{BarEve85}. Hence, dividing its result by a factor of two results in a lower bound on the size of the MVC, which is an upper bound on the size of MC.
            \item We employ the matrix rank upper bound on the size of the maximum independent set of \cite{Budinich2003}. Since a clique in $G$ is an independent set in $\overline{G}$, we obtain an upper bound on the clique size. Likewise, since the complement of any independent set of vertices is a vertex cover, any upper bound on the maximum independent set size corresponds to a lower bound on the MVC size.
            \item We use the easy to compute \textit{minimum degree bound} of \cite[page~20]{Willis2011}.
        \end{enumerate}
    We will refer to this bound as \textit{deterministic bound}.
    \item Any graph coloring provides an upper bound on the \textit{chromatic number}, or the minimum number of colors needed to color the vertices of a graph so that each edge connects vertices of different colors. Since all vertices in a maximum clique need to be assigned different colors, the chromatic number is again an upper bound of MC. This means that it gives an upper bound on the size of the maximum independent set of the complement graph and thus a lower bound on MVC. 
    We will refer to this bound as \textit{chromatic bound}. Computing the chromatic number is NP-hard, so its exact computation would be intractable, but there are much better heuristics for its approximation compared with the ones for e.g.\ the clique number. Therefore, a greedy search heuristic for the chromatic number provides an easily computable bound. To compute a graph coloring, we use the heuristic function \textit{greedy\_color} of the NetworkX package \citep{Networkx}, which is applied to the complement $\overline{G}$ of $G$.
\end{enumerate}
Analogously, the following are \textit{lower bounds} for MC and \textit{upper bounds} for MVC:
\begin{enumerate}
    \item For MC, $G^-$ is the larger of the two subgraphs and it contains $G^+$. (However, note that $G^+$ cannot simply be ignored since the MC number of $G$ is computed as $\omega(G)=\min(\omega(G^+)+1,\omega(G^-))$.)
    For MVC, $G^+$ is the larger subgraph. Thus at any point in the decomposition tree, the best solution found so far in $G^-$ ($G^+$) can be used to get a lower (upper) bound on the size of the MC (MVC). Therefore, we first follow the appropriate recursive branches in the decomposition until we can compute the MC or MVC on any of the generated subgraphs: its size can then be used to get a good lower (upper) bound for the size of MC (MVC) in all the other (smaller) generated subgraphs. We call this strategy the \textit{decomposition bound}.
    \item We apply the \textit{fmc} maximum clique solver \citep{fmc} in fast heuristic mode to $G$, thus giving us a lower bound on the size of MC. Since the size of the MVC of $G$ added to the clique number of $\overline{G}$ equals the size of $G$, finding an approximation of the clique number in $\overline{G}$ with the help of the \textit{fmc} heuristic translates to an upper bound on the size of the MVC of $G$. We will denote this strategy as the \textit{fmc bound}.
\end{enumerate}
The above bounds are employed during the recursion to prune those subproblems which cannot contain vertices belonging to the MC or MVC of the input graph.

\subsection{Reduction techniques for MC}
\label{sec:mc_reductions}
In addition to using upper and lower bounds, we also use two \textit{reduction techniques} that allow us to reduce the size of a subproblem during the decomposition. The first reduction works directly on the subgraphs, the second one works with the QUBO formulation of MC given in \eqref{eq:MC}.
\begin{enumerate}
    \item The \textit{(vertex) $k$-core algorithm} can reduce the number of vertices of the input graph in some cases, and the \textit{edge $k$-core algorithm} \citep{Chapuis2017, DBLP:journals/corr/cs-DS-0310049} can reduce the number of edges.

    The (vertex) $k$-core of a graph $G = (V, E)$ was defined in Section~\ref{sec:vertexchoice} as the maximal subgraph of $G$ in which every vertex has a degree of at least $k$. Therefore, if a graph has a clique $C$ of size $k+1$, then this clique $C$ must be contained in the $k$-core of $G$ and all vertices outside of the $k$-core can be removed.

    The \textit{edge $k$-core} of a graph $G$ is defined in \cite{Chapuis2017}. It is easily shown that for two vertices $v$, $w$ in a clique of size $c$, the intersection $N(v) \cap N(w)$ of the two neighbor lists $N(v)$ and $N(w)$ of $v$ and $w$ has size at least $c-2$. Denoting the current best lower bound on the clique size as $L$, we can therefore choose a random vertex $v$ and remove all edges $(v,w)$ satisfying $|N(v) \cap N(w)| < L-1$, since such edges cannot be part of a clique with size larger than $L$.
    \item Another reduction technique works on the QUBO formulation of MC given in \eqref{eq:MC}. In particular, for any subgraph produced by our algorithm, we generate the corresponding QUBO formulation of the MC problem \eqref{eq:MC}, which is then analyzed. Several general-purpose preprocessing techniques are capable to identify values of a subset of variables, called \textit{persistencies}, in a QUBO or Ising problem. Persistencies determine the value of certain variables in every global minimum (strong persistencies) or at least one global minimum (weak persistencies). In \cite{Boros2002}, a comprehensive overview of such techniques is given. Suppose variable $x_v$ for vertex $v$ (see Section~\ref{sec:intro}) is assigned the value $x_v=1$ in the persistency analysis: we can then add $v$ to the current set of vertices belonging to the maximum clique and remove $v$ and its adjacent edges from the subgraph. If $x_v=0$, we can remove $v$ and its adjacent edges without further processing. We employ the \textit{qpbo} Python bindings of \cite{Rother2007} to carry out the persistency analysis.
\end{enumerate}
We use the vertex and edge $k$-core algorithms with $k$ being set to the current best lower bound value for the clique number, thus allowing one to prune entire subgraphs that cannot contain a clique of size larger than the best current one.

\subsection{Reduction techniques for MVC}
\label{sec:mvc_reductions}
Similarly to Section~\ref{sec:mc_reductions}, we employ three reduction techniques to reduce the size of MVC instances during the decomposition. The first two are applied to the subgraphs, the last one works directly on the QUBO formulation of MVC given in \eqref{eq:MVC}.
\begin{enumerate}
    \item We coin the first method \textit{neighbor-based vertex removal} (abbreviated as \textit{nbvr}). Essentially, we search and remove triangles, vertices of degree one, and vertices of degree zero in any subgraph. Since in a triangle, any two arbitrarily chosen degree two triangle vertices belong to the MVC, we add a contribution of two to the overall size of the MVC and remove the triangle. Analogously, vertices of degree one are automatically in the MVC and can be removed, along with the only neighbor of that vertex, after adding a contribution of one to the current size of the MVC. Vertices of degree zero can be removed without further processing.
    \item In \cite{AkibaIwata2015}, the authors state a variety of reduction techniques for MVC that have been used in the theoretical study of exponential-complexity branch-and-reduce algorithms. For instance, those techniques include \textit{degree-one reductions}, \textit{decomposition}, \textit{dominance rules}, \textit{unconfined vertex reduction}, \textit{LP-} and \textit{packing reductions}, as well as \textit{folding-}, \textit{twin-}, \textit{funnel-} and \textit{desk reductions}. We employ only the reduction methods from the Java package \textit{vertex\_cover-master} of \cite{javaVC}. For a given input graph, \textit{vertex\_cover-master} returns a superset of the MVC. This means that any vertex not contained in the output of \textit{vertex\_cover-master} is definitely not part of the MVC and can be removed. The code can be applied repeatedly until no further vertices are found that can be removed.
    \item As for MC, we compute the QUBO formulation for MVC given by \eqref{eq:MVC} for any generated subgraph and apply persistency analysis to it using the \textit{qpbo} Python implementation of \cite{Rother2007}. Resolved variables during the analysis, e.g., those with $x_v=1$ for any $v \in V$, indicate that a particular vertex belongs to the minimum vertex cover. We can then add that vertex $v$ to the current optimal solution and remove it from the subgraph together with its adjacent edges. Analogously we remove those $v \in V$ with $x_v=0$.
\end{enumerate}
The simulations in Section~\ref{sec:experiments} assess the effectiveness of the aforementioned bounds (Section~\ref{sec:experiments_bounds}) and reduction techniques (Section~\ref{sec:experiments_reductions}).

\section{Experimental results}
\label{sec:experiments}
In the experiments we look at various aspects of the proposed decomposition method in Algorithm~\ref{algo:decomposition}. We start with an assessment of the vertex choice of Section~\ref{sec:vertexchoice}. We then evaluate the proposed bounds (Section~\ref{sec:bounds}) and reduction techniques (Sections~\ref{sec:mc_reductions} and \ref{sec:mvc_reductions}). Using the best combination of vertex choice, bounding and reduction techniques, we concretize the algorithms for MC (Section~\ref{sec:MC}) and MVC (Section~\ref{sec:MVC}) and evaluate them in Section~\ref{sec:experiments_algo}. An application of our algorithms to real world graphs is presented in Section~\ref{sec:realworld}. A prediction regarding their scaling behavior on future D-Wave architectures is presented in Section~\ref{sec:experiments_future}.

Throughout the simulations we use three measures to assess the performance in all experiments:
\begin{enumerate}
    \item subgraph count: The total number of subgraphs produced during the decomposition.
    \item preprocessing CPU time: The total time for the decomposition alone without actual solving of any subproblems.
    \item predicted solution time: This time accounts for the total preprocessing time and factors in an average QPU access time of $1.6$ seconds for $10000$ anneals on D-Wave 2000Q for each of the generated subgraphs on leaf level. Thus the predicted solution time is estimated using the formula: subgraph count $\times$ $1.6$ seconds + preprocessing time. Note that the solutions returned by D-Wave 2000Q after $10000$ anneals might not always be optimal.
\end{enumerate}
In all experiments apart from Section~\ref{sec:experiments_future}, we always run the decomposition until the subgraphs produced in the recursion reach $64$ vertices, the largest size of an arbitrary problem that can embedded onto the D-Wave 2000Q hardware. As test graphs, we employ Erd\H{o}s–R\'{e}nyi random graphs \citep{ErdosRenyi1960} with $100$ vertices and an edge density ranging from $0.1$ to $0.9$ in steps of $0.1$.

\subsection{Evaluation of the vertex selection}
\label{sec:experiments_vertexchoice}
\begin{figure}
\begin{minipage}{0.49\textwidth}
    \centering
    \includegraphics[width=\textwidth]{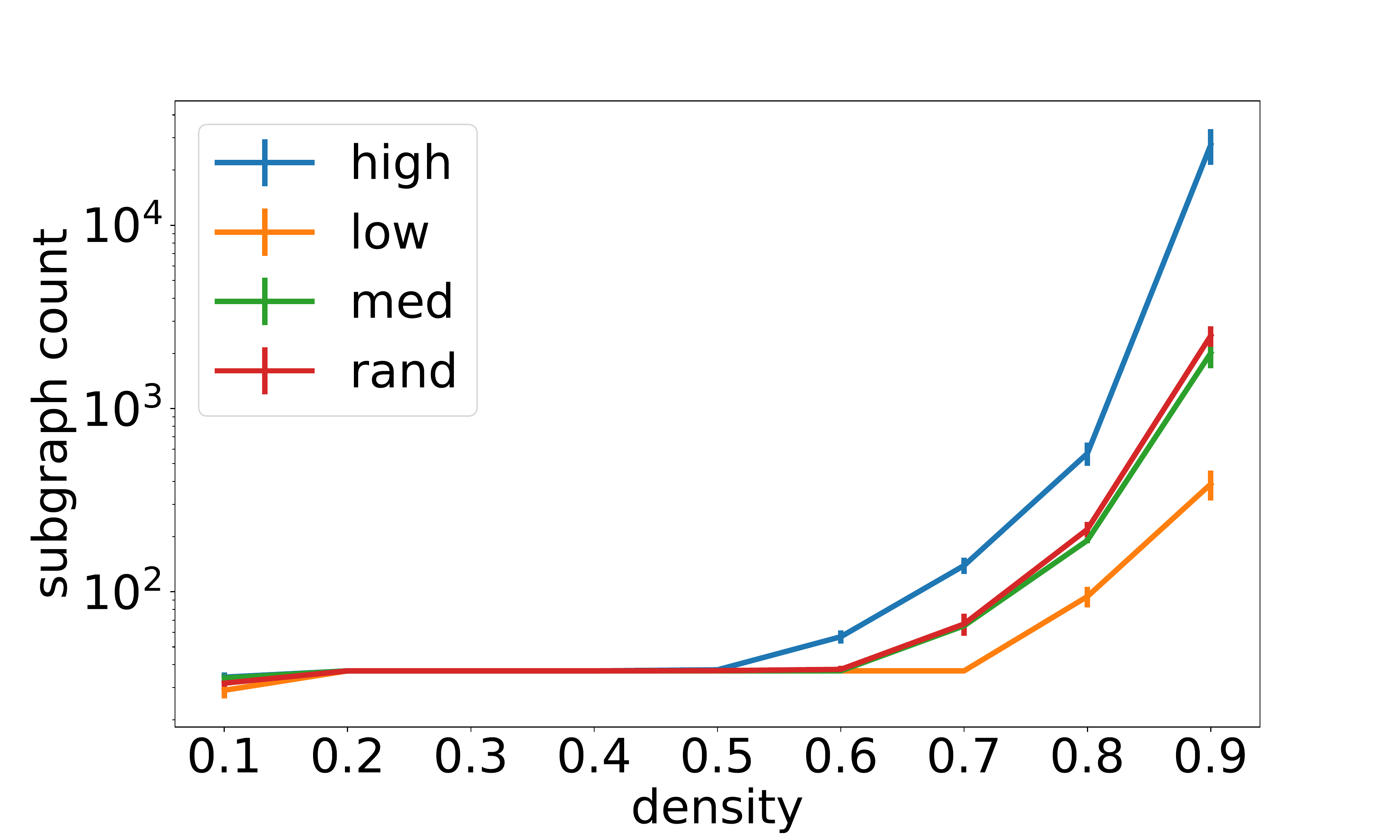}\\
    \includegraphics[width=\textwidth]{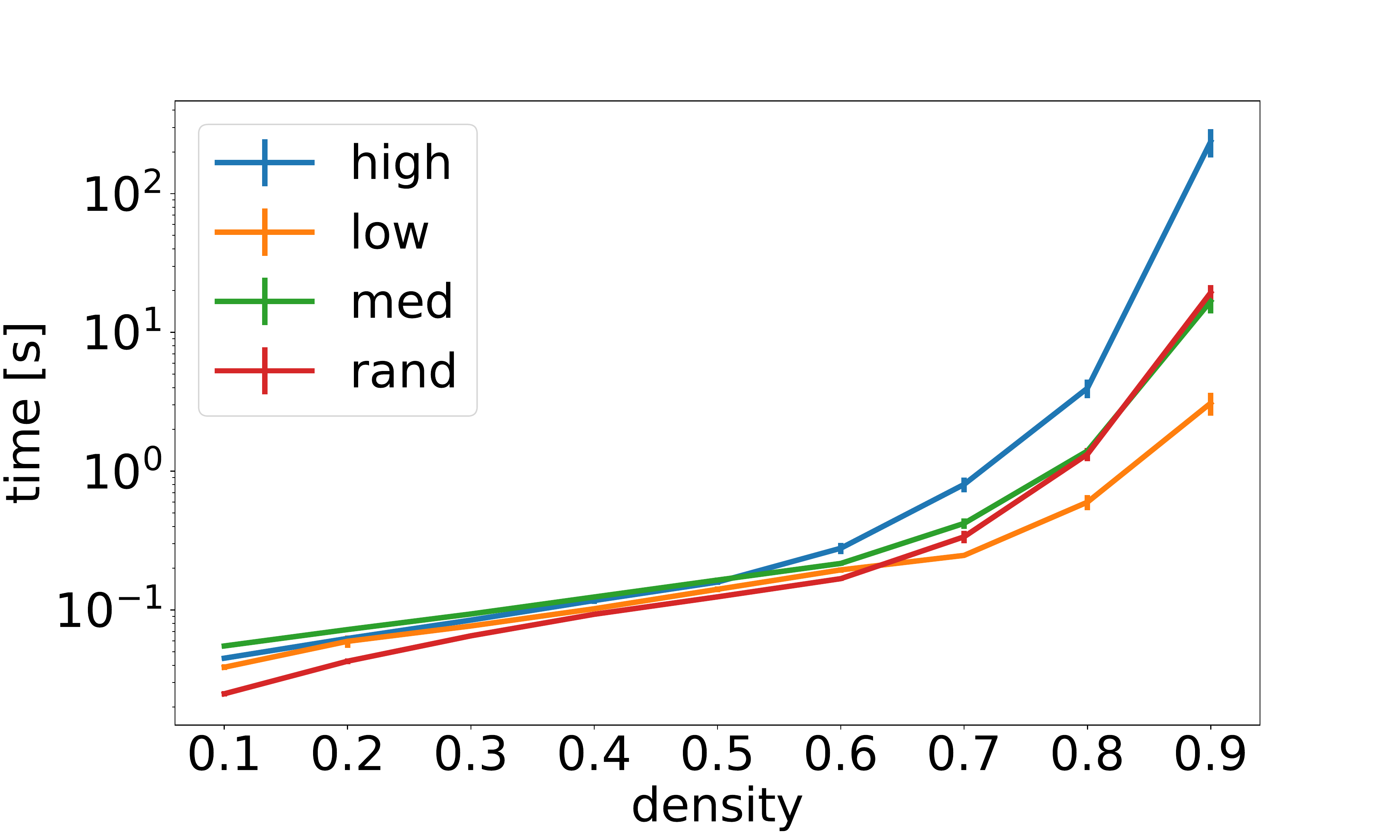}\\
    \includegraphics[width=\textwidth]{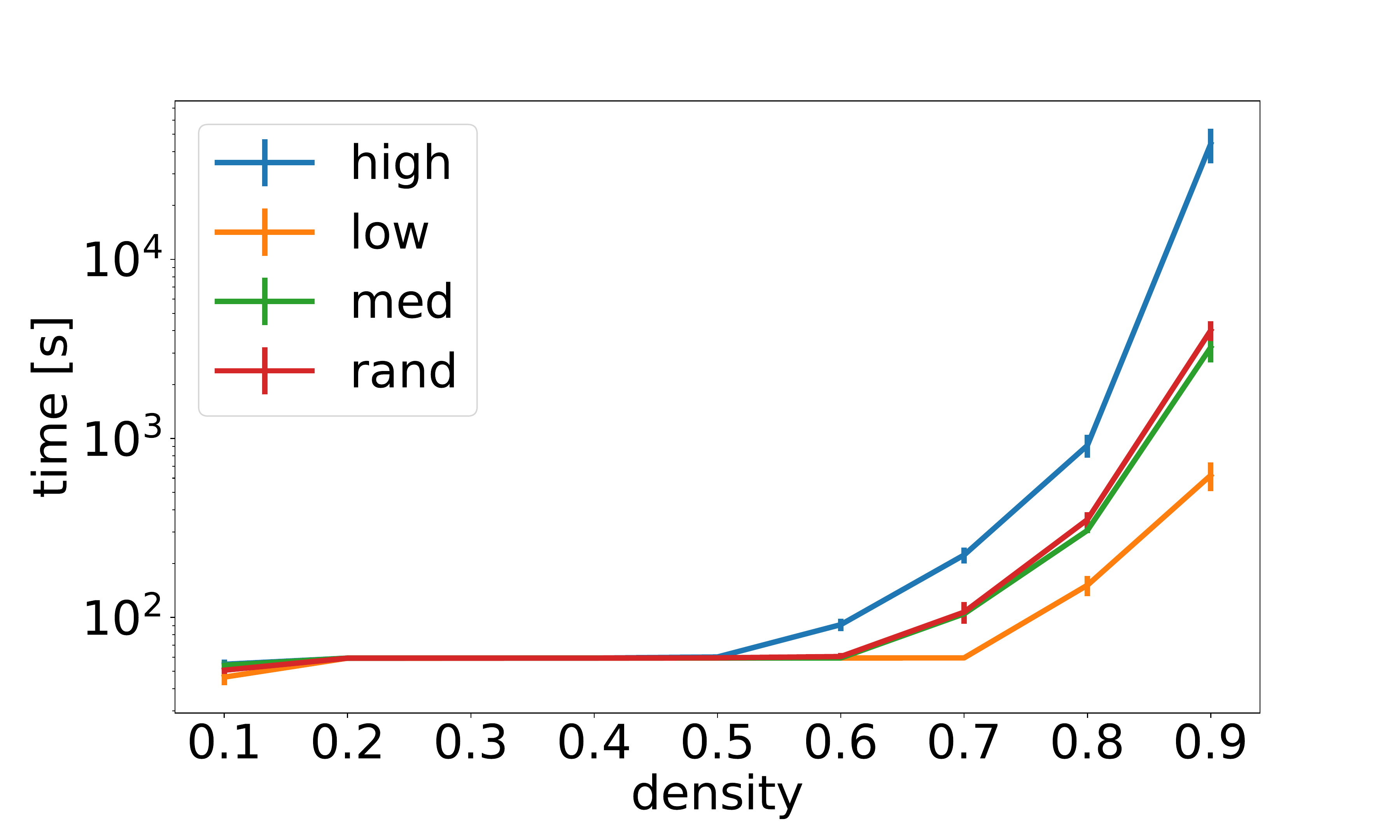}
    \caption{Vertex selection strategies for MC: high, low, median degree vertex and random vertex. Subgraph count (top), preprocessing time (middle) and predicted solution time (bottom) as a function of the graph density. Logarithmic scale on the y-axis. Error bars of one standard deviation.}
    \label{fig:mc_vertex_choice}
\end{minipage}\hfill
\begin{minipage}{0.49\textwidth}
    \centering
    \includegraphics[width=\textwidth]{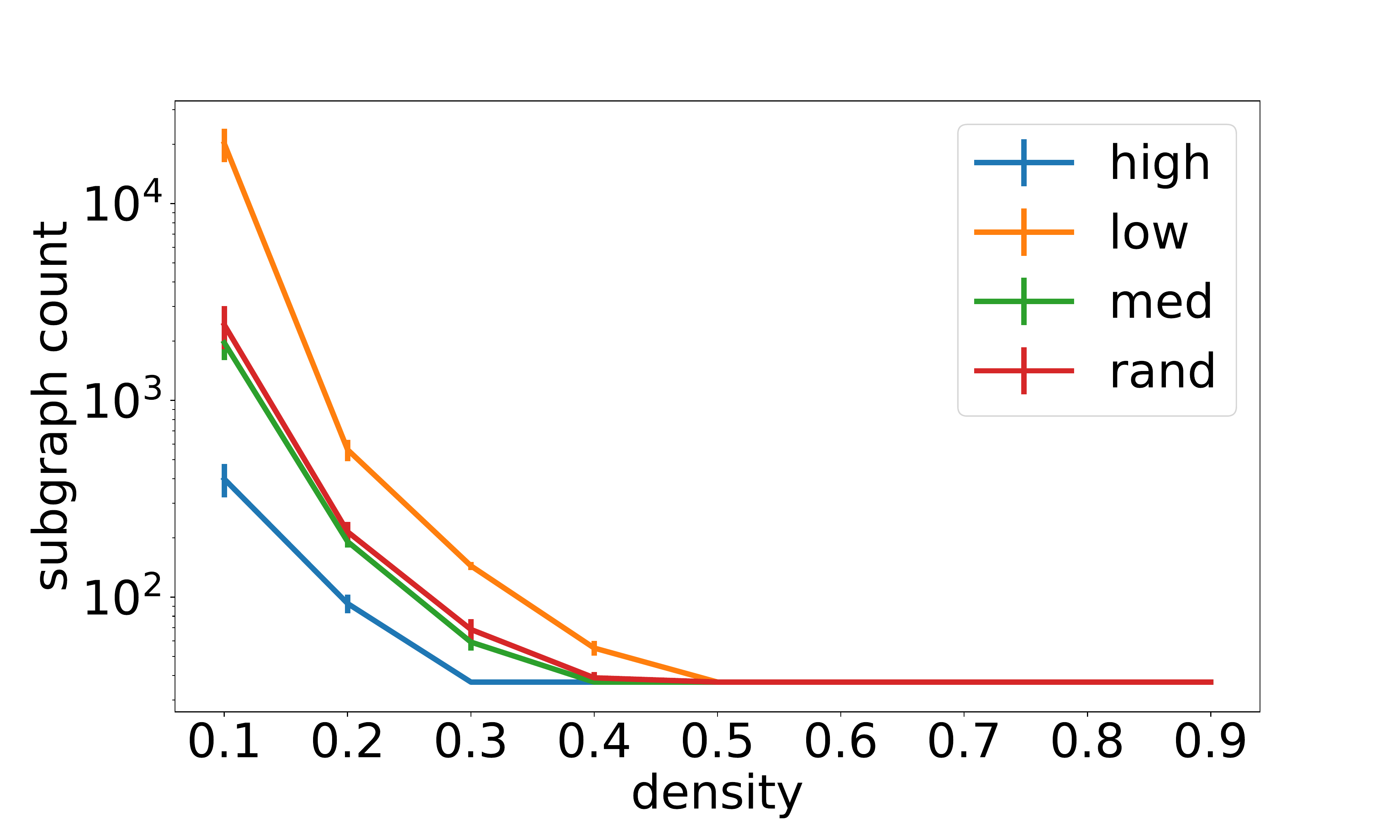}\\
    \includegraphics[width=\textwidth]{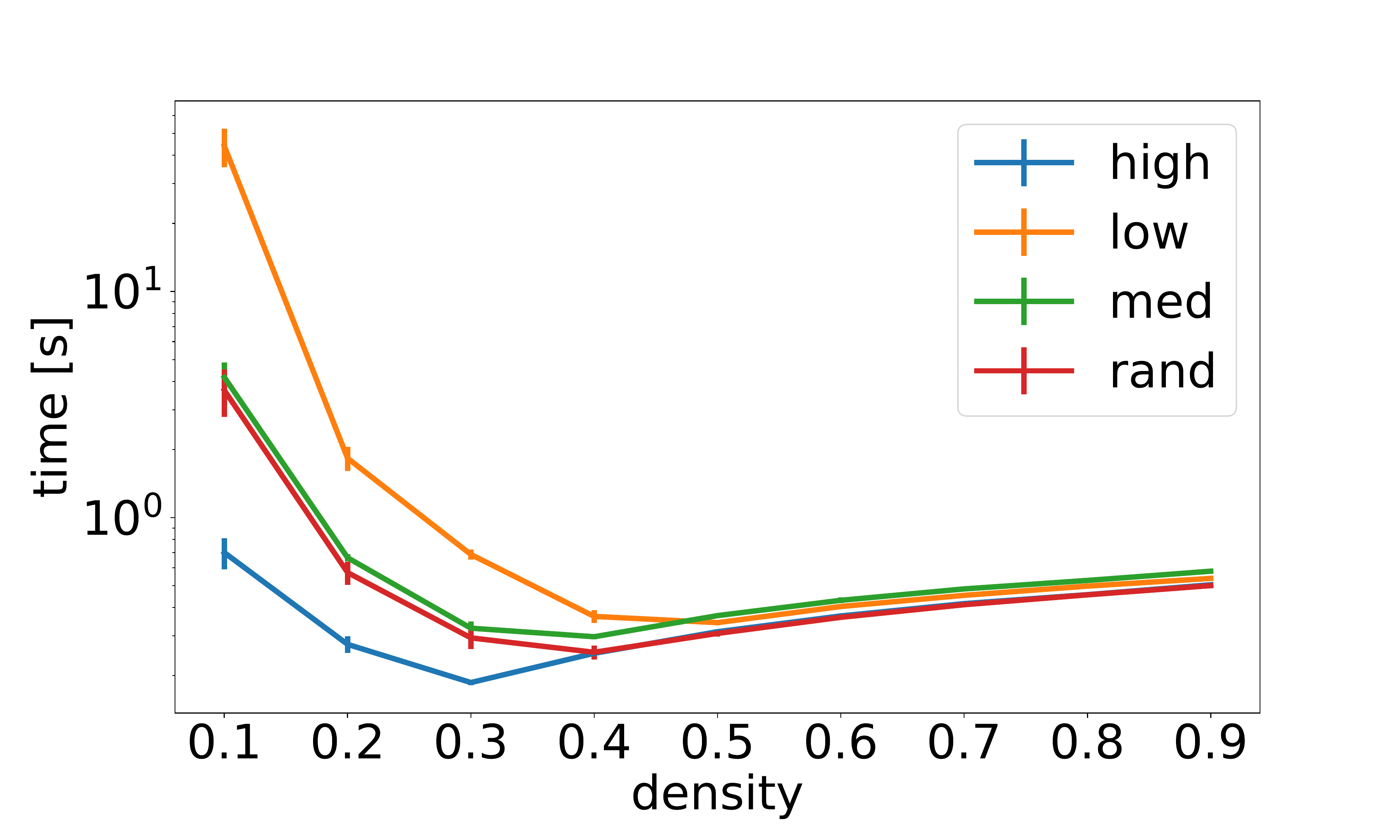}\\
    \includegraphics[width=\textwidth]{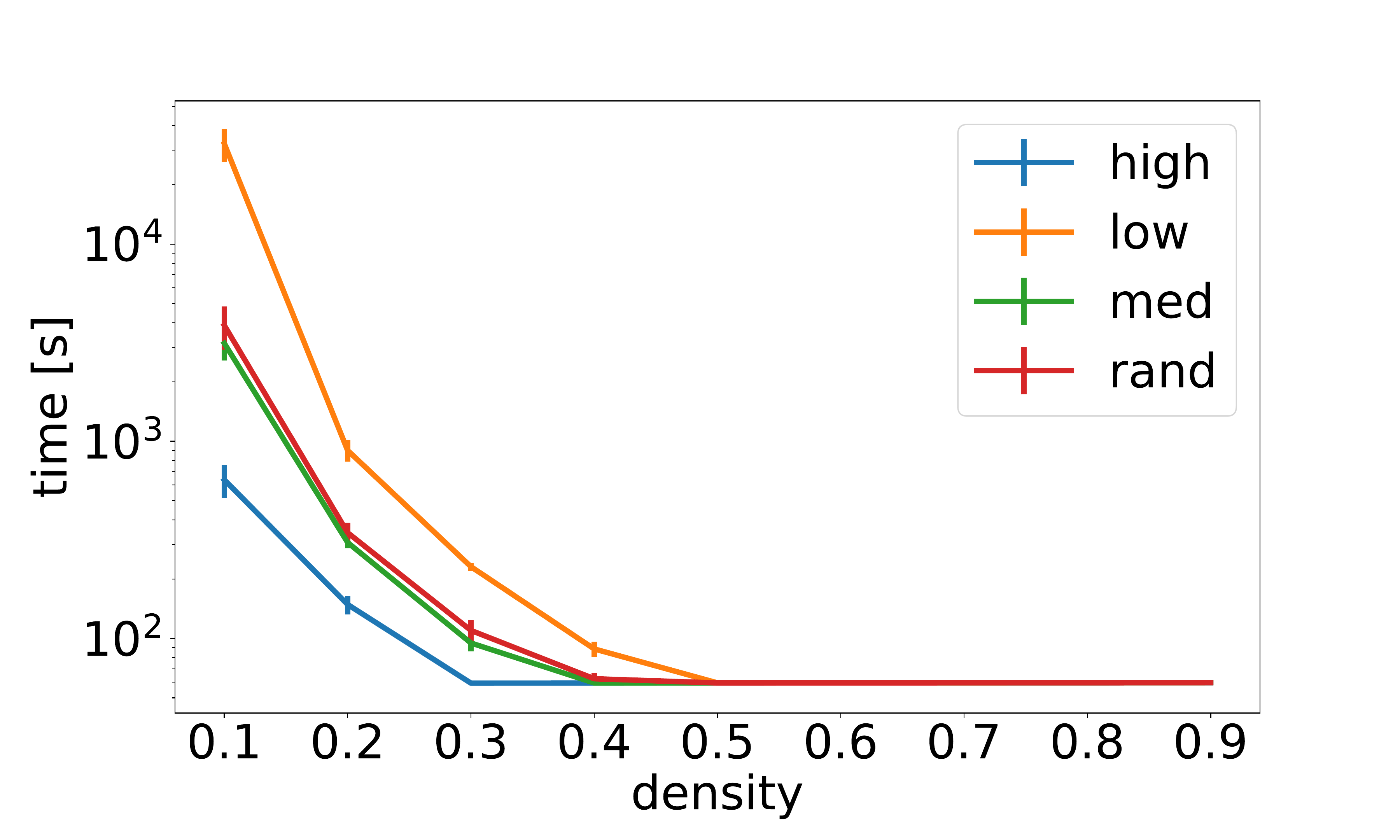}
    \caption{Vertex selection strategies for MVC: high, low, median degree vertex and random vertex. Subgraph count (top), preprocessing time (middle) and predicted solution time (bottom) as a function of the graph density. Logarithmic scale on the y-axis. Error bars of one standard deviation.}
    \label{fig:mvc_vertex_choice}
\end{minipage}
\end{figure}

We start with an assessment of the strategies for vertex selection discussed in Section~\ref{sec:vertexchoice}. Figure~\ref{fig:mc_vertex_choice} presents results for subgraph count, preprocessing and predicted solution times for MC and the four vertex selection choices of Section~\ref{sec:vertexchoice}. We observe that the four strategies roughly agree for low densities across all measures, though for very low densities the random vertex selection has a slight advantage in terms of subgraph count. However, for densities above $0.5$, selecting a lowest degree vertex yields best performance across all measures, while a highest degree vertex selection performs worst.

For MVC, which is the inverse problem of MC, we also observe an inverted picture in Figure~\ref{fig:mvc_vertex_choice}. Here, selecting a highest degree vertex yields best performance across all measures for low graph densities, while lowest degree vertex selection performs worst. The intuition here is that, if the vertex degree is $k$, then the graph $G^-$ will have $k-1$ fewer vertices then $G$. For higher densities, the four strategies yield roughly similar results.

\subsection{Lower and upper bounds}
\label{sec:experiments_bounds}
\begin{figure}
\begin{minipage}{0.49\textwidth}
    \centering
    \includegraphics[width=\textwidth]{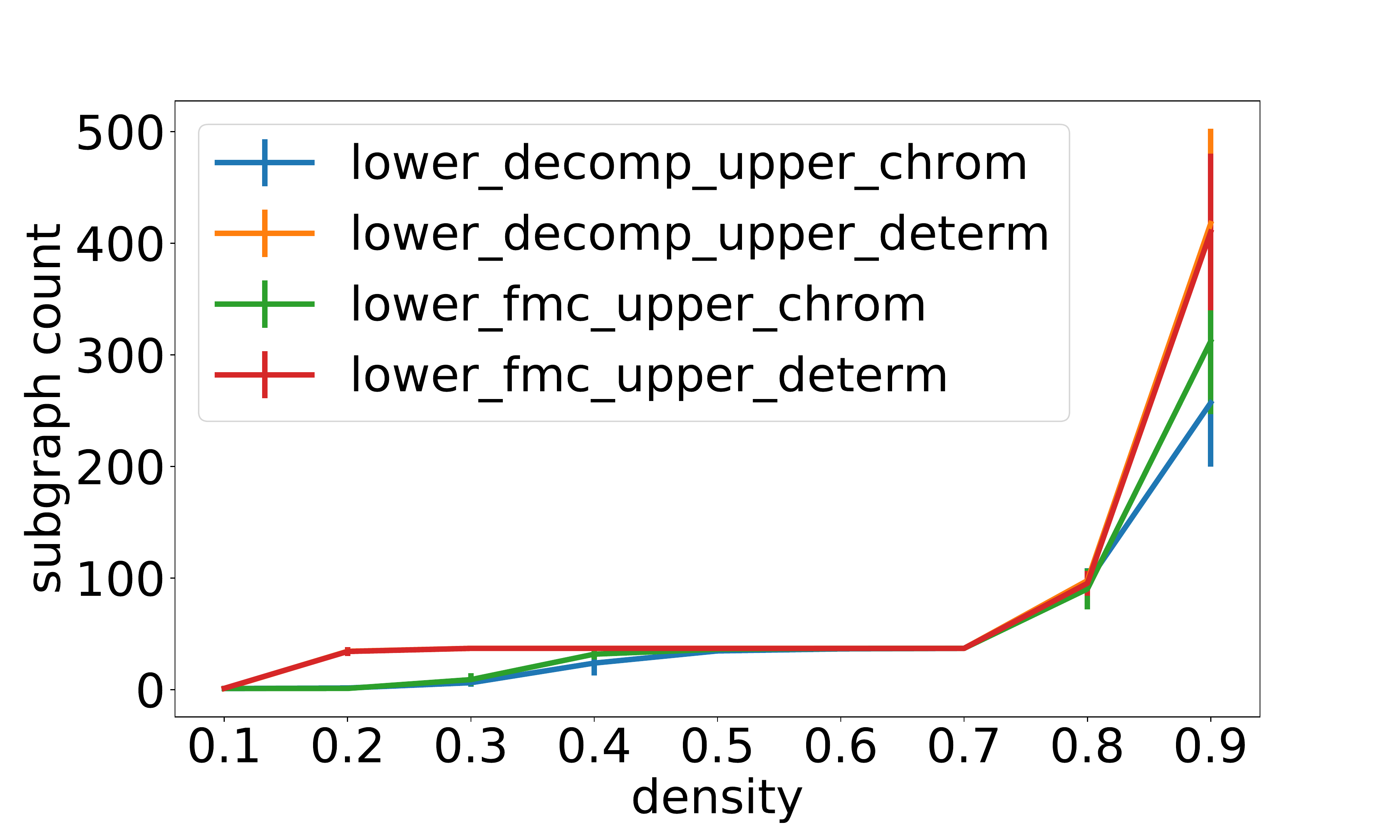}\\
    \includegraphics[width=\textwidth]{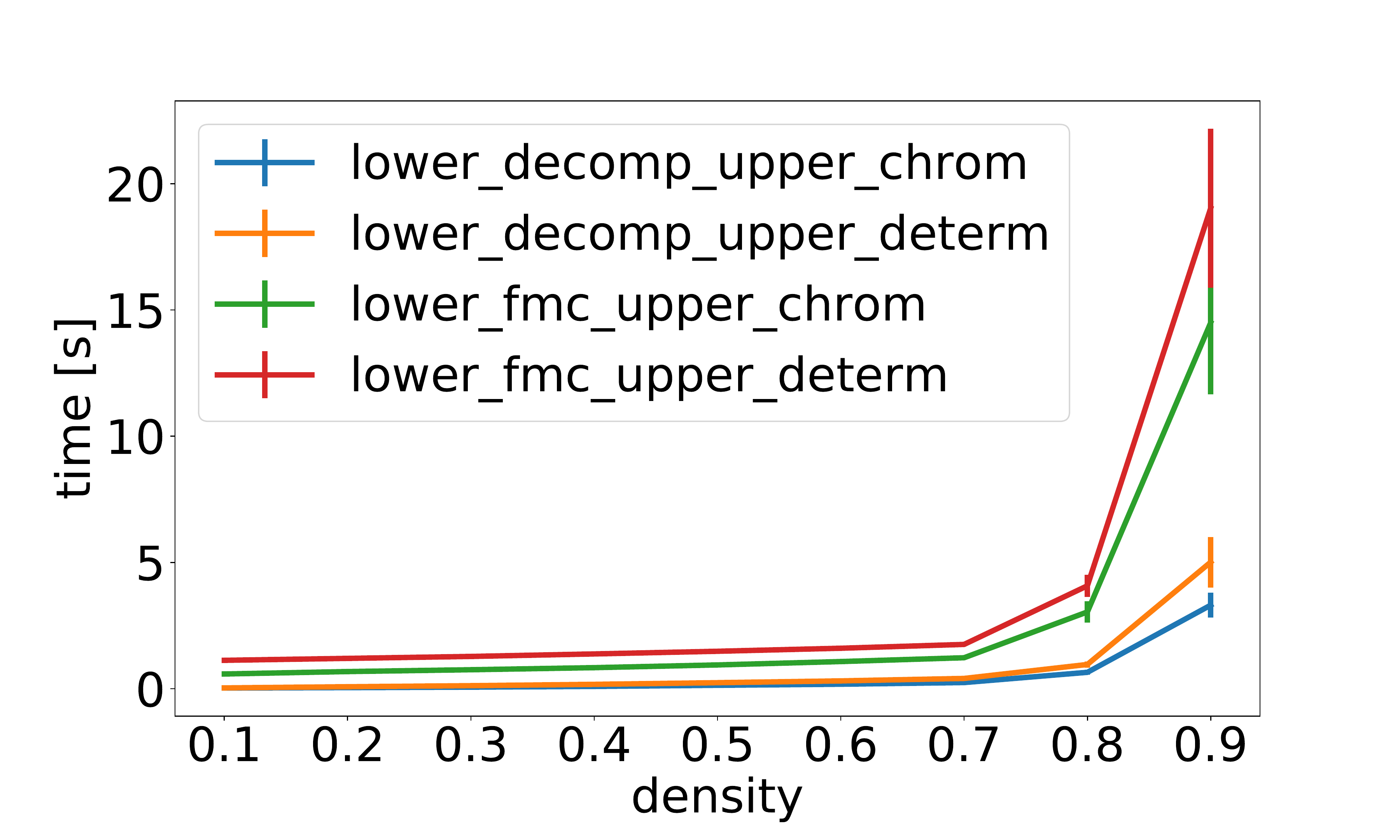}\\
    \includegraphics[width=\textwidth]{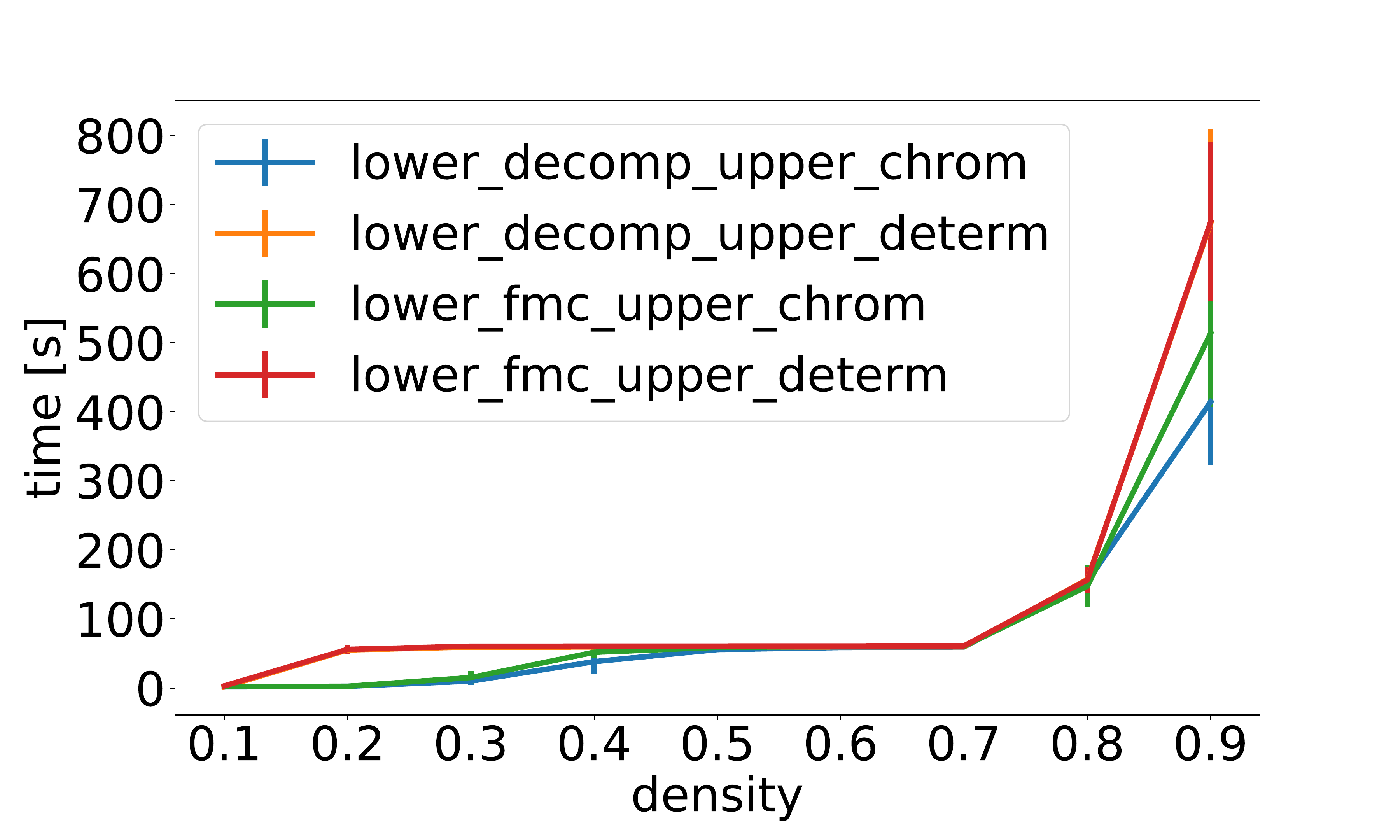}
    \caption{All combinations of lower and upper bounds for MC. Lower bounds are the decomposition and fmc bounds, upper bounds are the chromatic and deterministic bounds (see Section~\ref{sec:bounds}). Subgraph count (top), preprocessing time (middle) and predicted solution time (bottom) as a function of the graph density. Error bars of one standard deviation.}
    \label{fig:mc_bounds}
\end{minipage}\hfill
\begin{minipage}{0.49\textwidth}
    \centering
    \includegraphics[width=\textwidth]{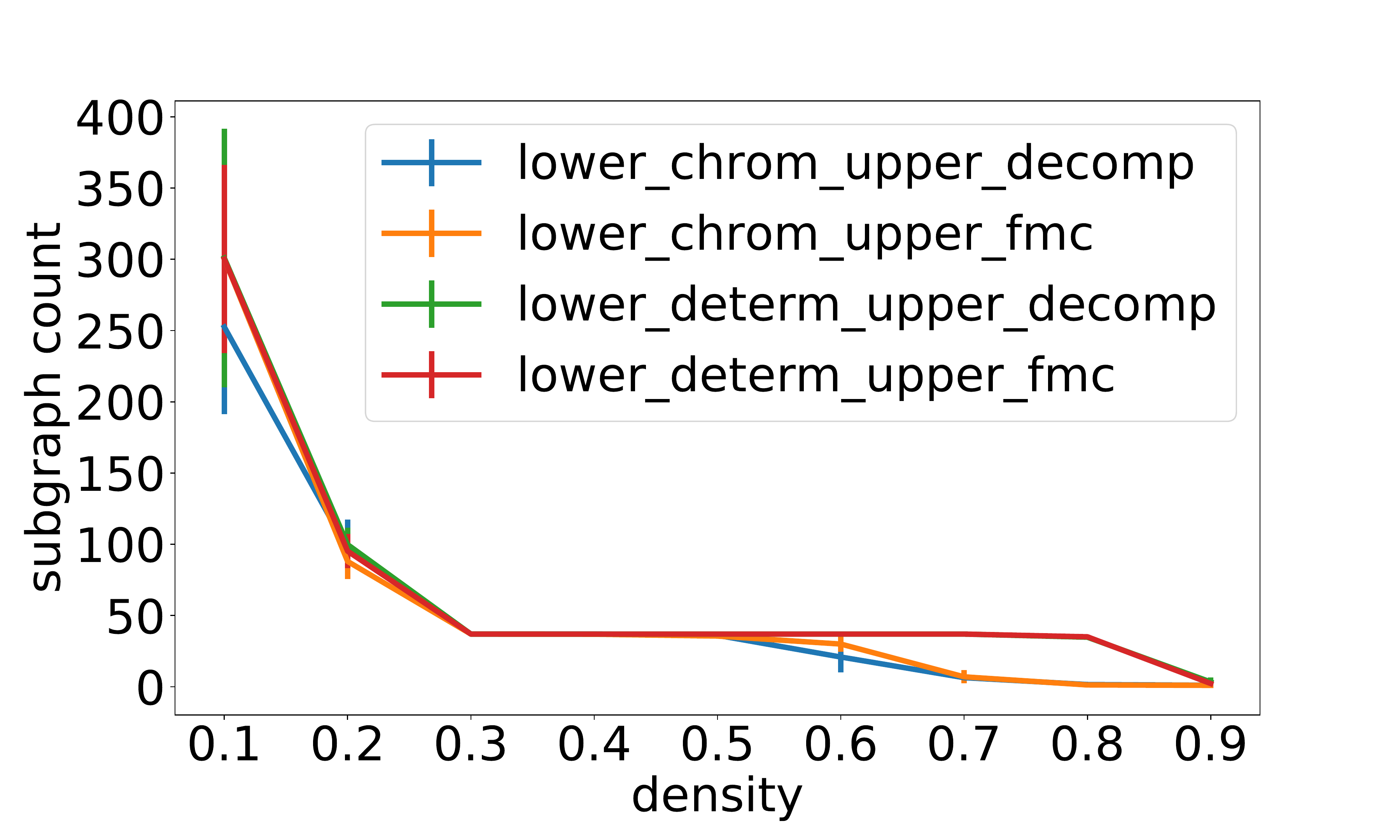}\\
    \includegraphics[width=\textwidth]{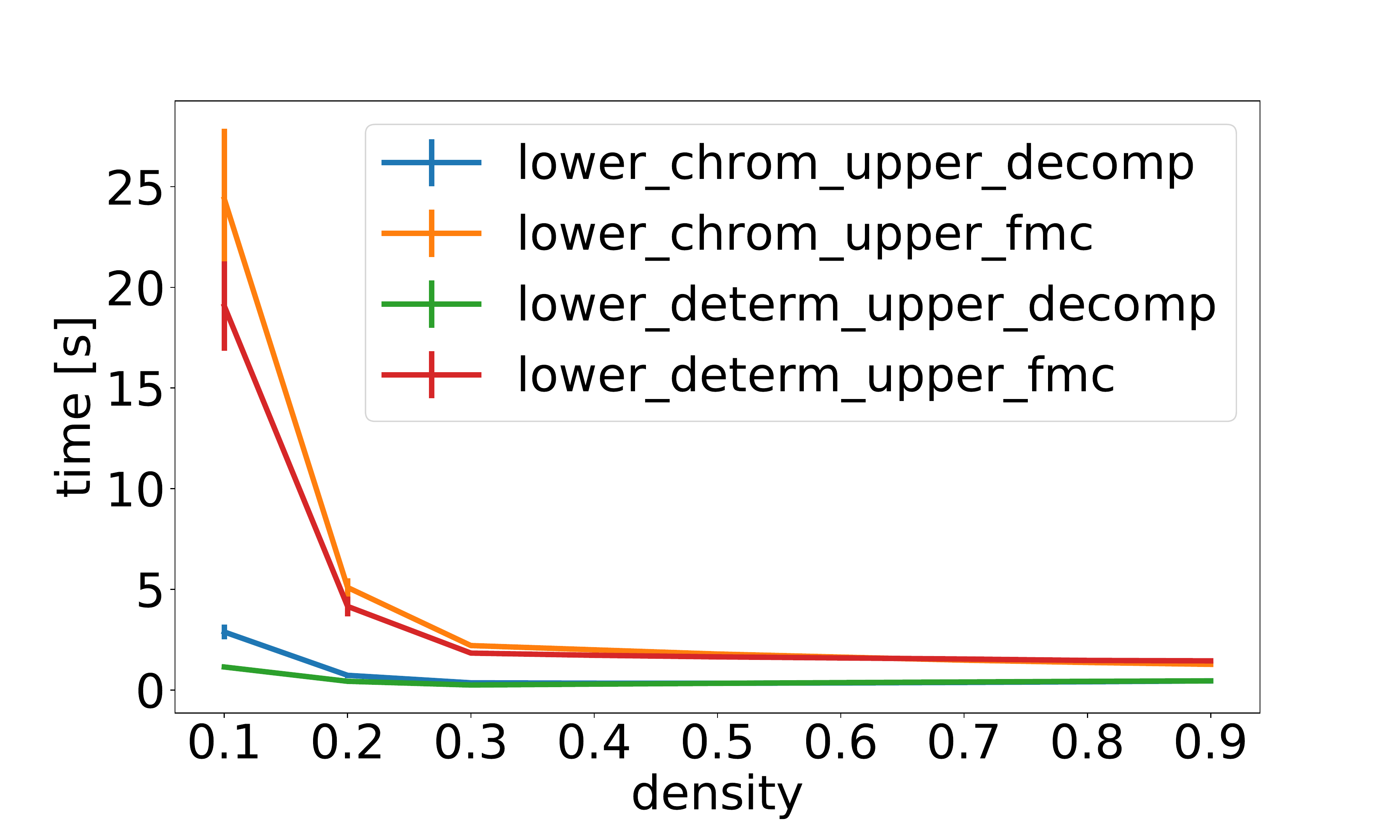}\\
    \includegraphics[width=\textwidth]{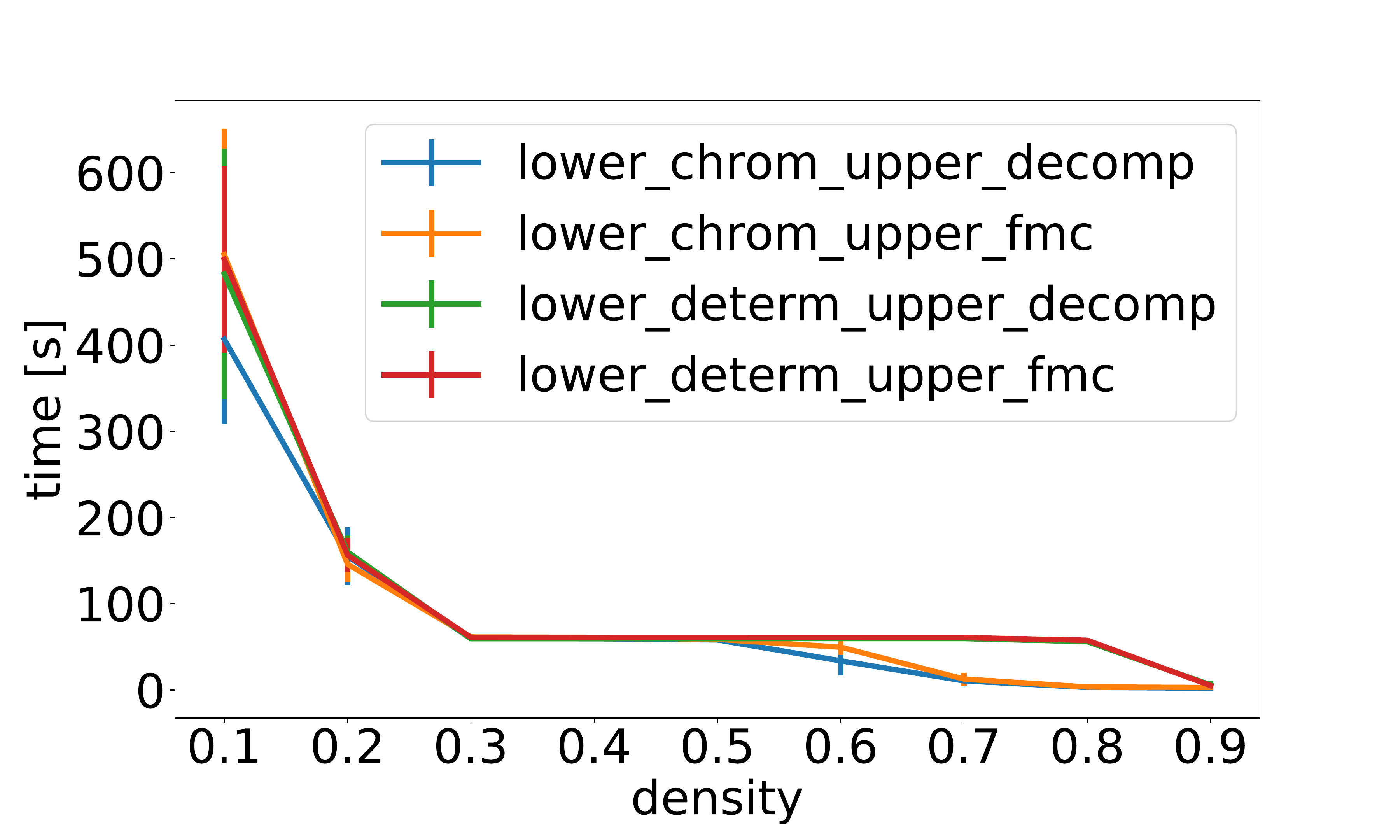}
    \caption{All combinations of lower and upper bounds for MVC. Lower bounds are the chromatic and deterministic bounds, upper bounds are the decomposition and fmc bounds (see Section~\ref{sec:bounds}). Subgraph count (top), preprocessing time (middle) and predicted solution time (bottom) as a function of the graph density. Error bars of one standard deviation.}
    \label{fig:mvc_bounds}
\end{minipage}
\end{figure}

We now evaluate the lower and upper bounds discussed in Section~\ref{sec:bounds}. For MC, the decomposition and fmc bounds are lower bounds, while the chromatic and deterministic bounds are upper bounds. Results are shown in Figure~\ref{fig:mc_bounds}. We observe that all combinations of bounds yield high reductions for low and moderate densities, and that the bounds become less effective for high densities. Of the four combinations tested, using the decomposition lower bound and the chromatic upper bound seems most advantageous as it yields the lowest subgraph count, is the quickest to compute, and results in an overall fastest runtime.

Figure~\ref{fig:mvc_bounds} repeats the assessment of the lower and upper bounds for MVC. Now, lower bounds for MVC are the chromatic and deterministic bounds, while the decomposition and fmc bounds are upper bounds. Since MC and MVC are inverse problems, the bounds are now less effective for low graph densities, and become highly effective for moderate and high densities. We again observe that overall, the combination of lower chromatic bound and upper decomposition bound is most advantageous.

\subsection{Reduction strategies}
\label{sec:experiments_reductions}
\begin{figure}
\begin{minipage}{0.49\textwidth}
    \centering
    \includegraphics[width=\textwidth]{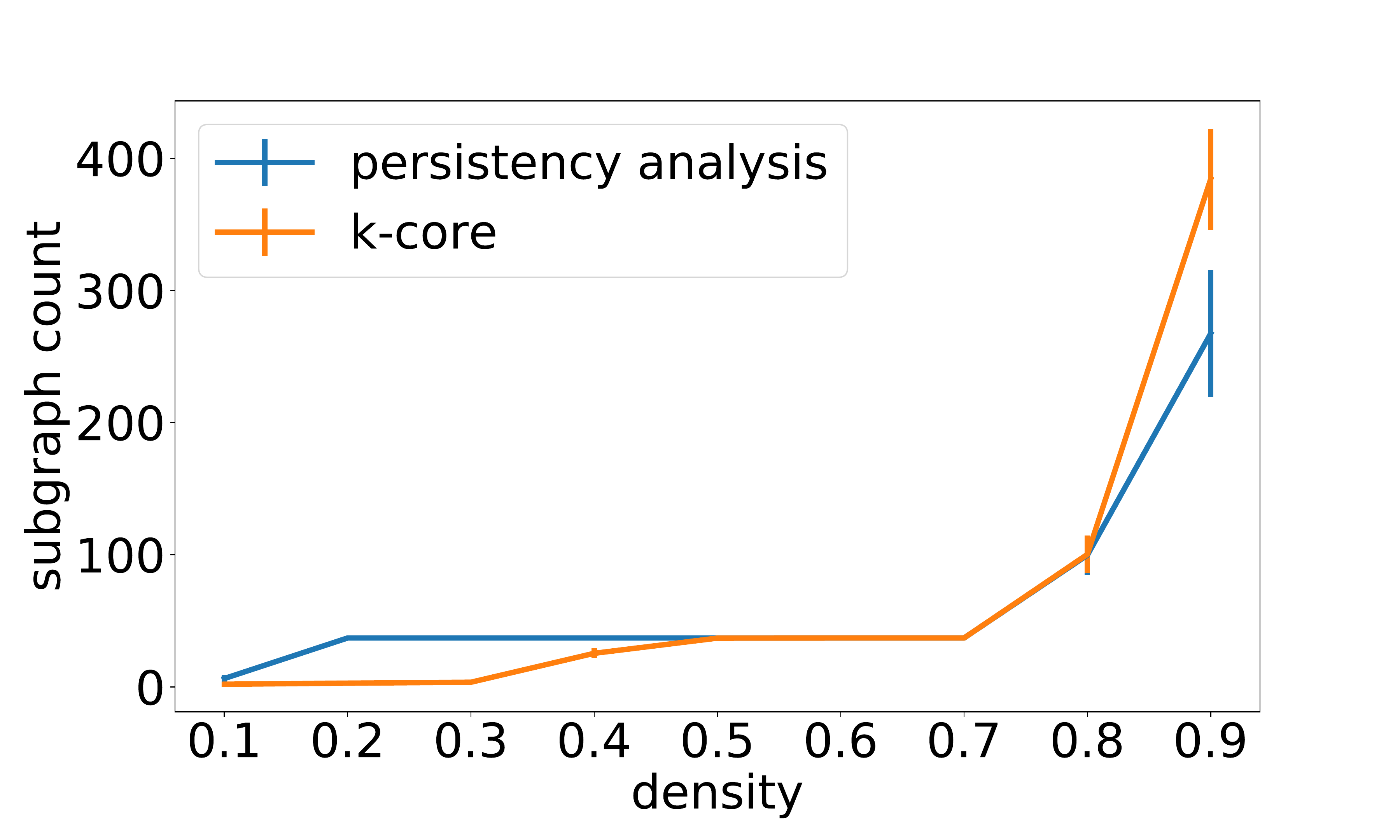}\\
    \includegraphics[width=\textwidth]{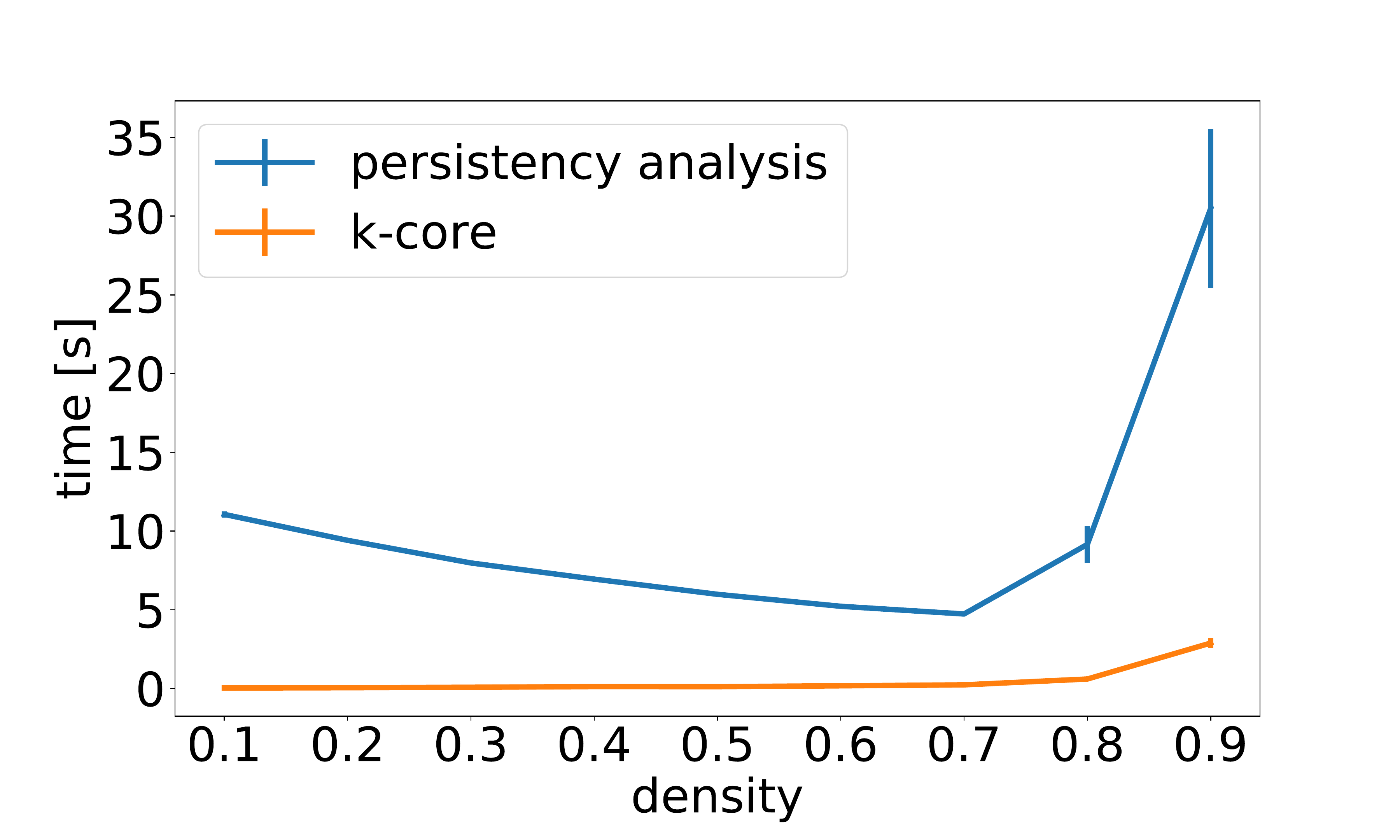}\\
    \includegraphics[width=\textwidth]{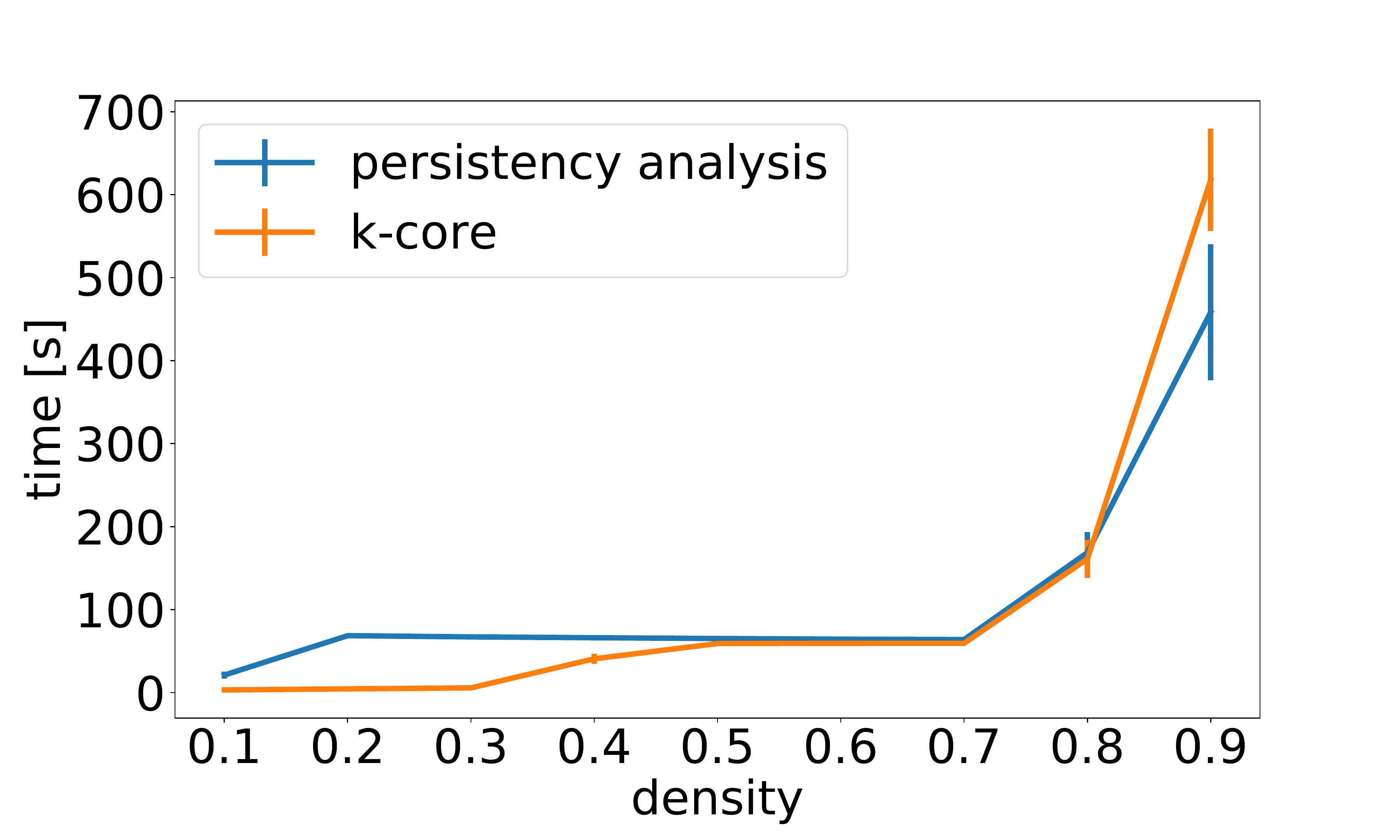}
    \caption{Persistency analysis and $k$-core reduction techniques for MC. Subgraph count (top), preprocessing time (middle) and predicted solution time (bottom) as a function of the graph density. Error bars of one standard deviation.}
    \label{fig:mc_reductions}
\end{minipage}\hfill
\begin{minipage}{0.49\textwidth}
    \centering
    \includegraphics[width=\textwidth]{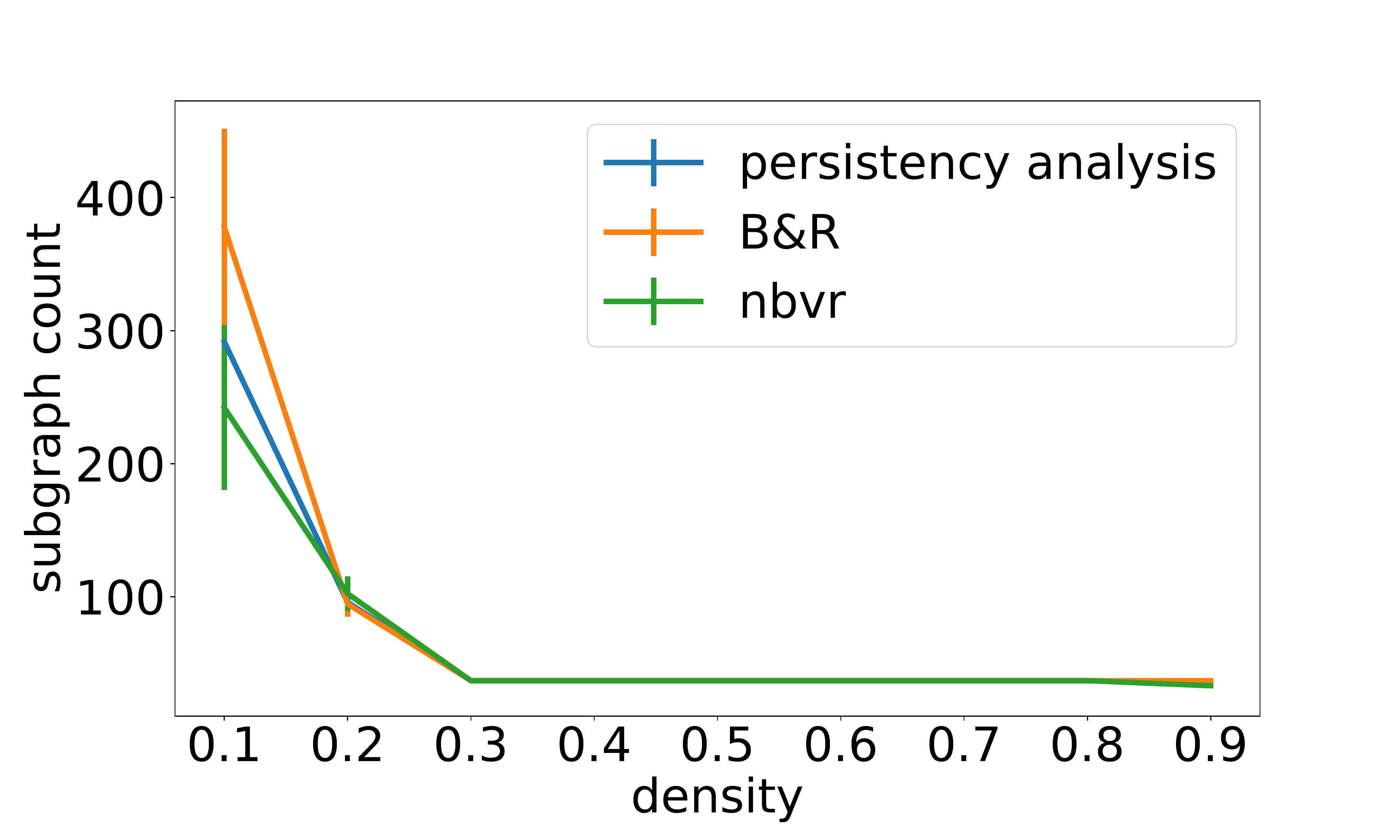}\\
    \includegraphics[width=\textwidth]{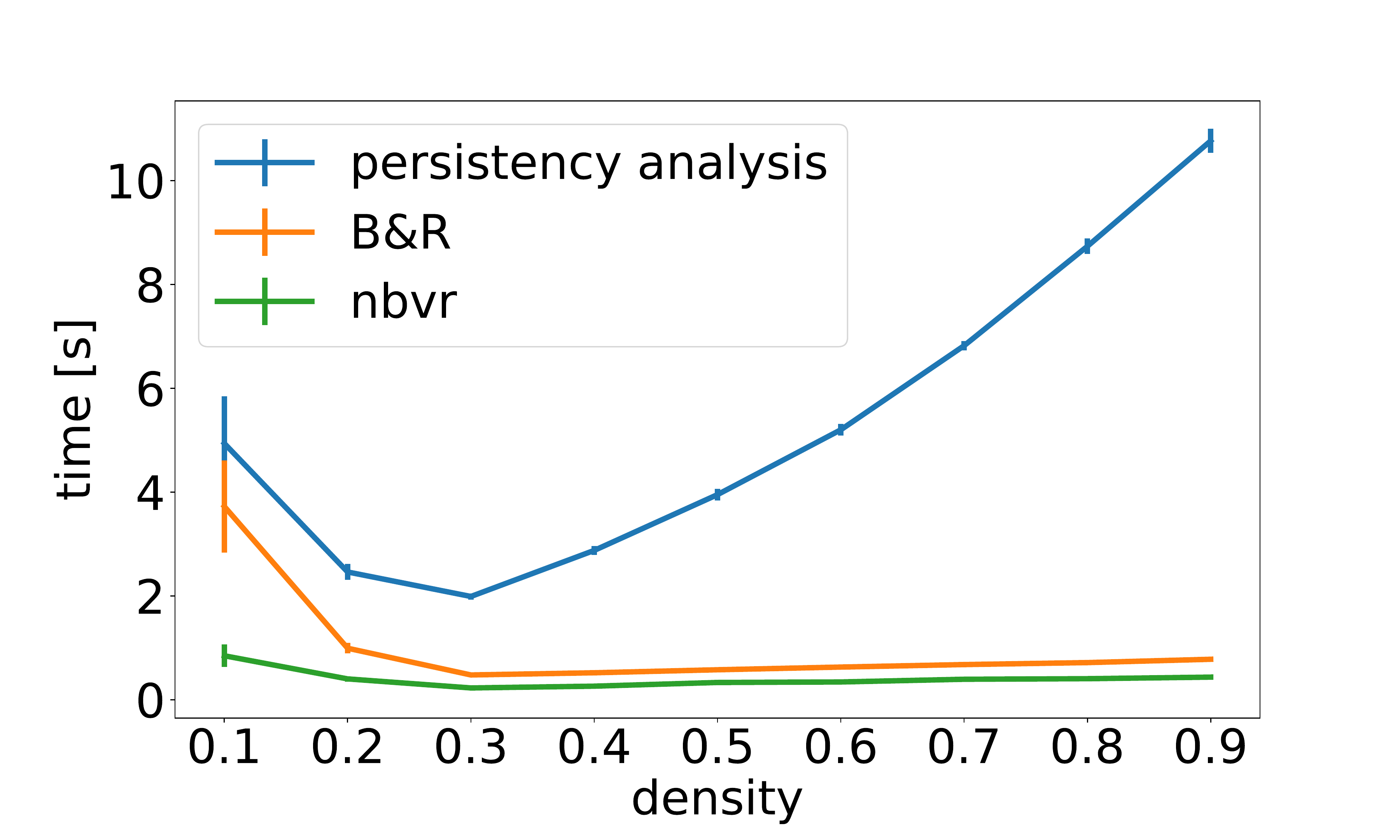}\\
    \includegraphics[width=\textwidth]{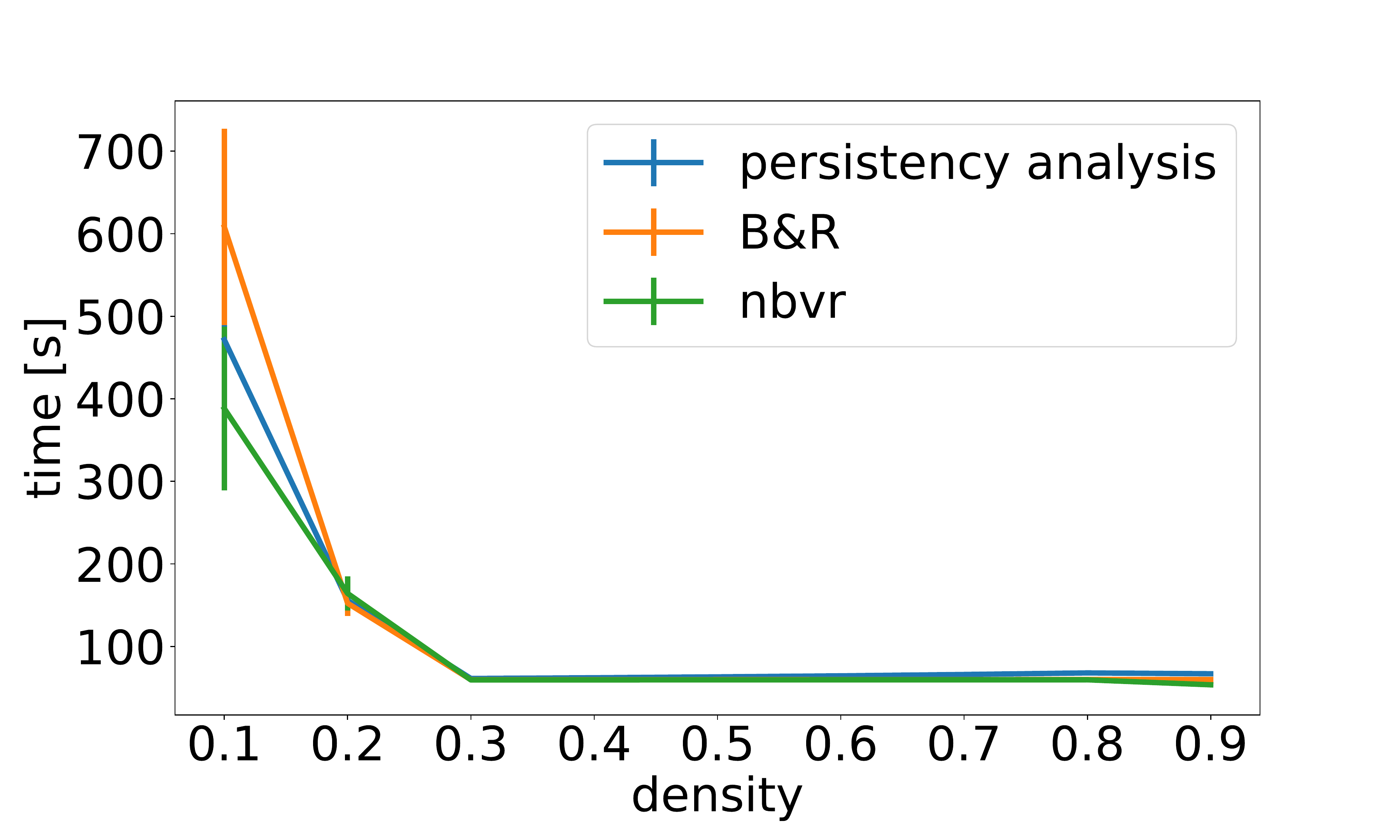}
    \caption{Persistency analysis, B\&R reduction and neighbor based vertex removal (nbvr) reduction techniques for MC. Subgraph count (top), preprocessing time (middle) and predicted solution time (bottom) as a function of the graph density. Error bars of one standard deviation.}
    \label{fig:mvc_reductions}
\end{minipage}
\end{figure}

Figure~\ref{fig:mc_reductions} assesses the behavior of the two reduction techniques for MC discussed in Section~\ref{sec:mc_reductions}. Those are the qpbo reduction which analyses the QUBO representation of MC given in \eqref{eq:MC}, and the $k$-core reduction which works on the graph itself.

We observe a particular behavior of these reductions in Figure~\ref{fig:mc_reductions}: overall, reductions are highly effective for low and moderate graph densities, and become less effective for high densities. While for low densities, $k$-core is better than qpbo, both techniques draw equal for moderate densities and qpbo overtakes $k$-core for high densities. The fact that qpbo becomes more effective for high densities has already been observed in \cite{Hahn2017ReducingBQ}. As shown in the plot depicting the preprocessing time, $k$-core is computationally efficient for all densities, while the computational complexity of qpbo is more complex but generally increases for high densities. The behavior of the predicted solution time again reflects the behavior of the subgraph count, with $k$-core being faster for low and medium densities, and slower for high densities.

Figure~\ref{fig:mvc_reductions} shows a similar comparison for the three reduction techniques for MVC outlined in Section~\ref{sec:mvc_reductions}. Those are qpbo working on the QUBO representation of MVC in \eqref{eq:MVC}, the reduction techniques of \cite{AkibaIwata2015} given in the Java package \textit{vertex\_cover-master} of \cite{javaVC}, and the neighbor based vertex removal (nbvr).

We observe that all reductions are able to reduce the subproblems most effectively for medium and high densities. Neighbor based vertex removal is fastest to compute, and persistency analysis again becomes slower with increasing graph density. Overall, neighbor based vertex removal is fastest to compute, yields the lowest number of subgraphs during decomposition and the lowest overall runtime.

\subsection{The DBK and DBR algorithms}
\label{sec:experiments_algo}
Using our preparatory experiments of Sections~\ref{sec:experiments_vertexchoice} to \ref{sec:experiments_reductions}, we can now fully specify the decomposition algorithm for MC introduced in Section~\ref{sec:MC}. To be precise, we employ the algorithm of Section~\ref{sec:MC} with low degree vertex selection (Section~\ref{sec:experiments_vertexchoice}), the decomposition lower and chromatic upper bounds (Section~\ref{sec:experiments_bounds}), and the $k$-core reduction strategy as determined in Section~\ref{sec:experiments_reductions}. As in \cite{Pelofske2019}, we call the resulting algorithm the DBK algorithm (Decomposition, Bounds, $K$-core).

\begin{figure}
\begin{minipage}{0.49\textwidth}
    \centering
    \includegraphics[width=\textwidth]{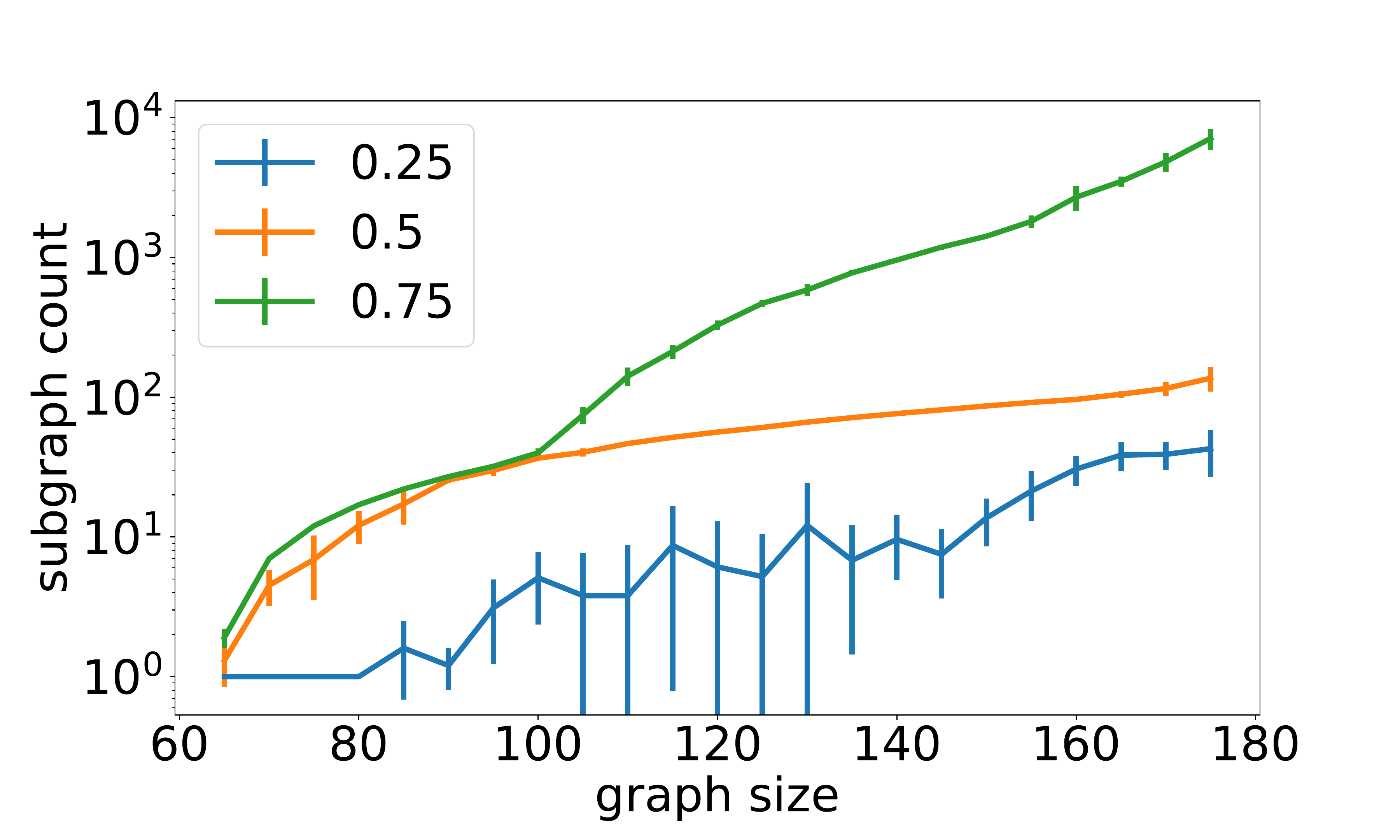}\\
    \includegraphics[width=\textwidth]{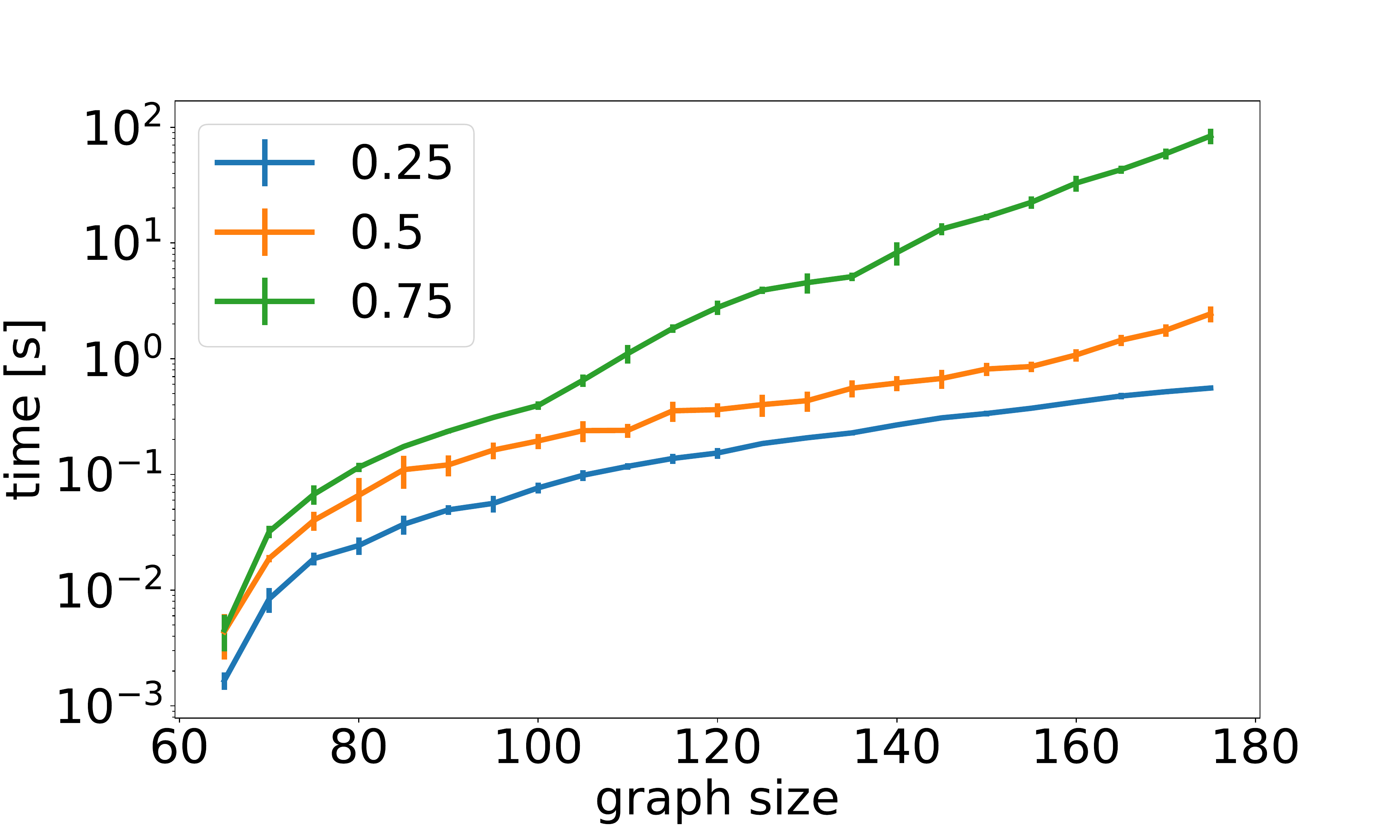}\\
    \includegraphics[width=\textwidth]{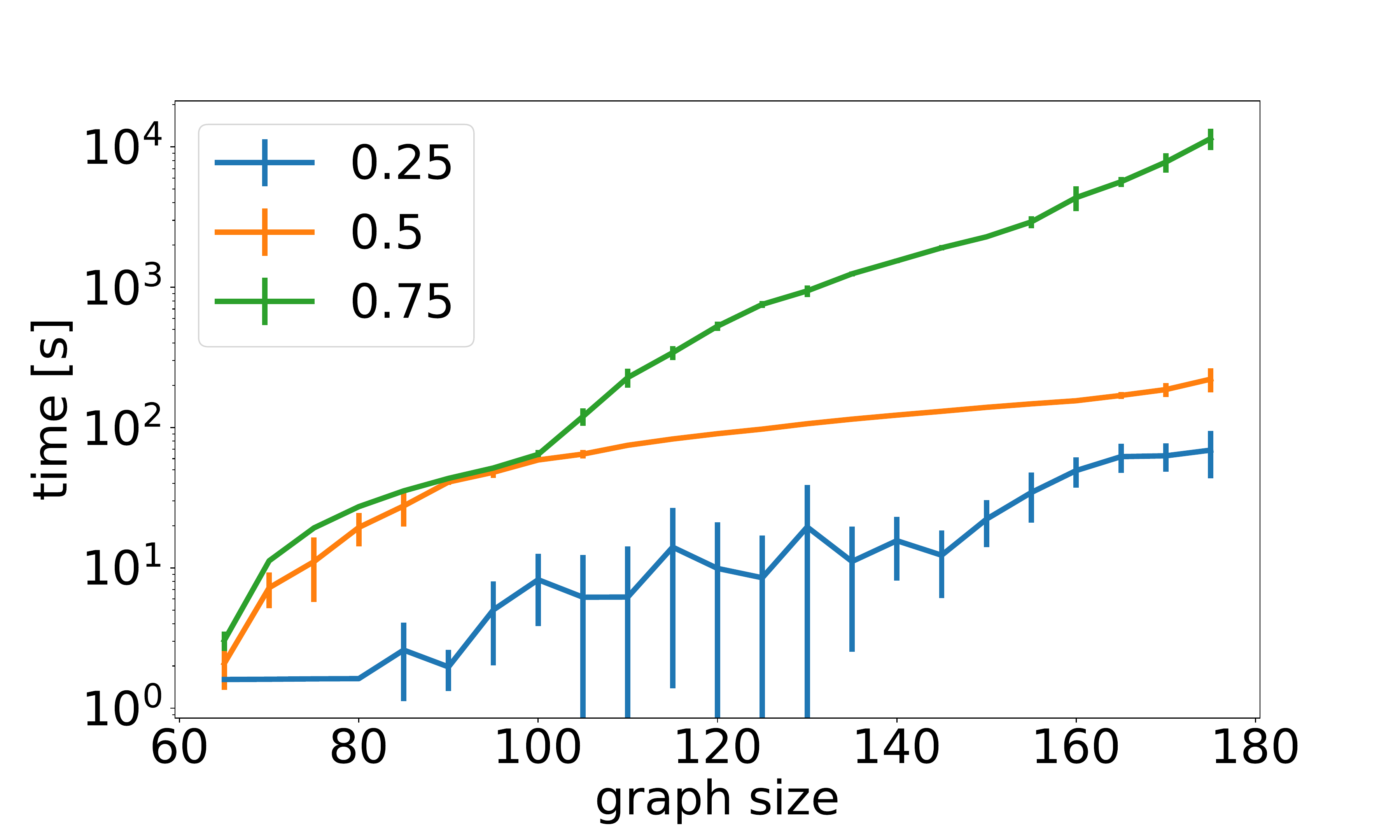}
    \caption{Performance of the DBK algorithm as a function of the graph size. Graph densities $d \in \{0.25,0.5,0.75\}$. Subgraph count (top), preprocessing time (middle) and predicted solution time (bottom) as a function of the graph density. Logarithmic scale on the y-axis. Error bars of one standard deviation.}
    \label{fig:dbk_scaling}
\end{minipage}\hfill
\begin{minipage}{0.49\textwidth}
    \centering
    \includegraphics[width=\textwidth]{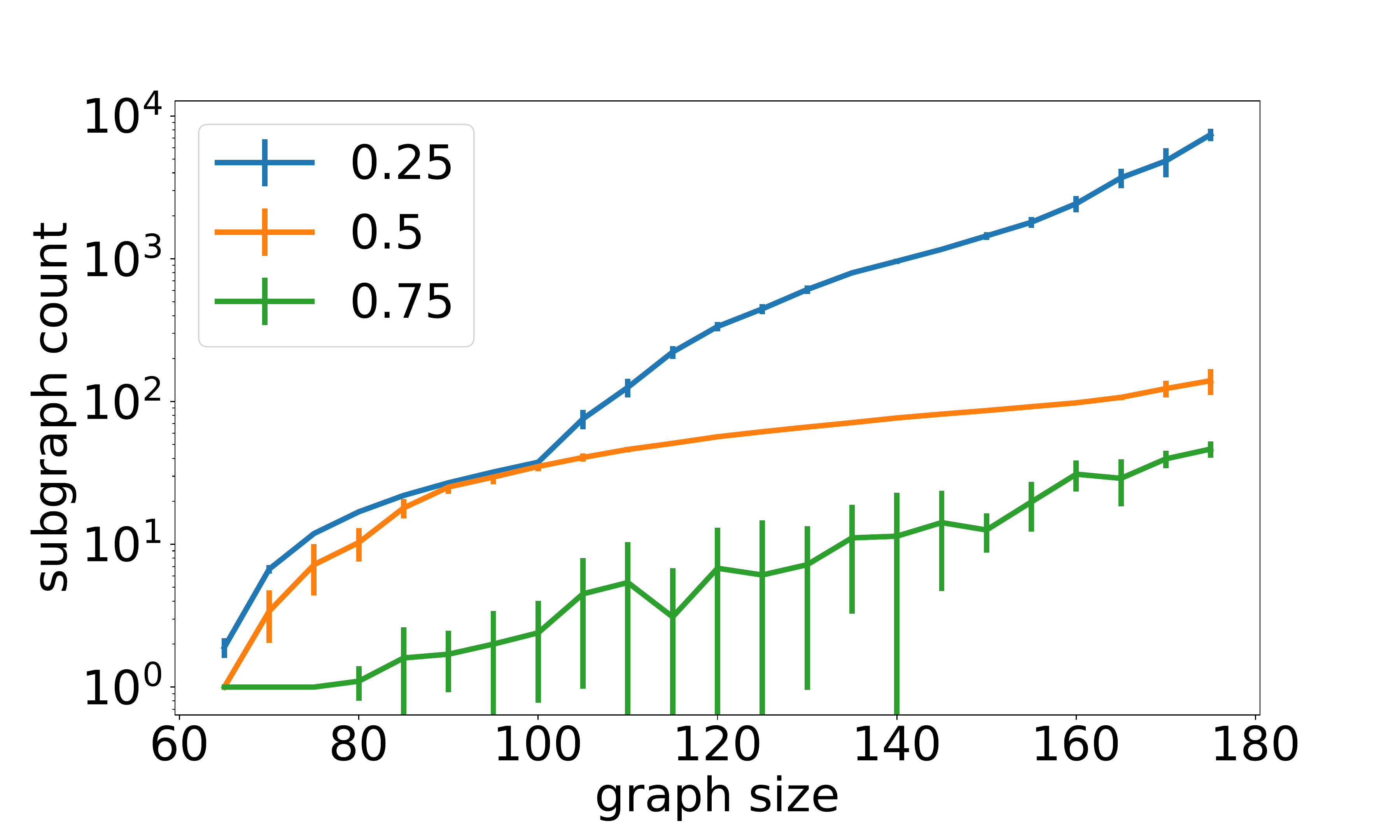}\\
    \includegraphics[width=\textwidth]{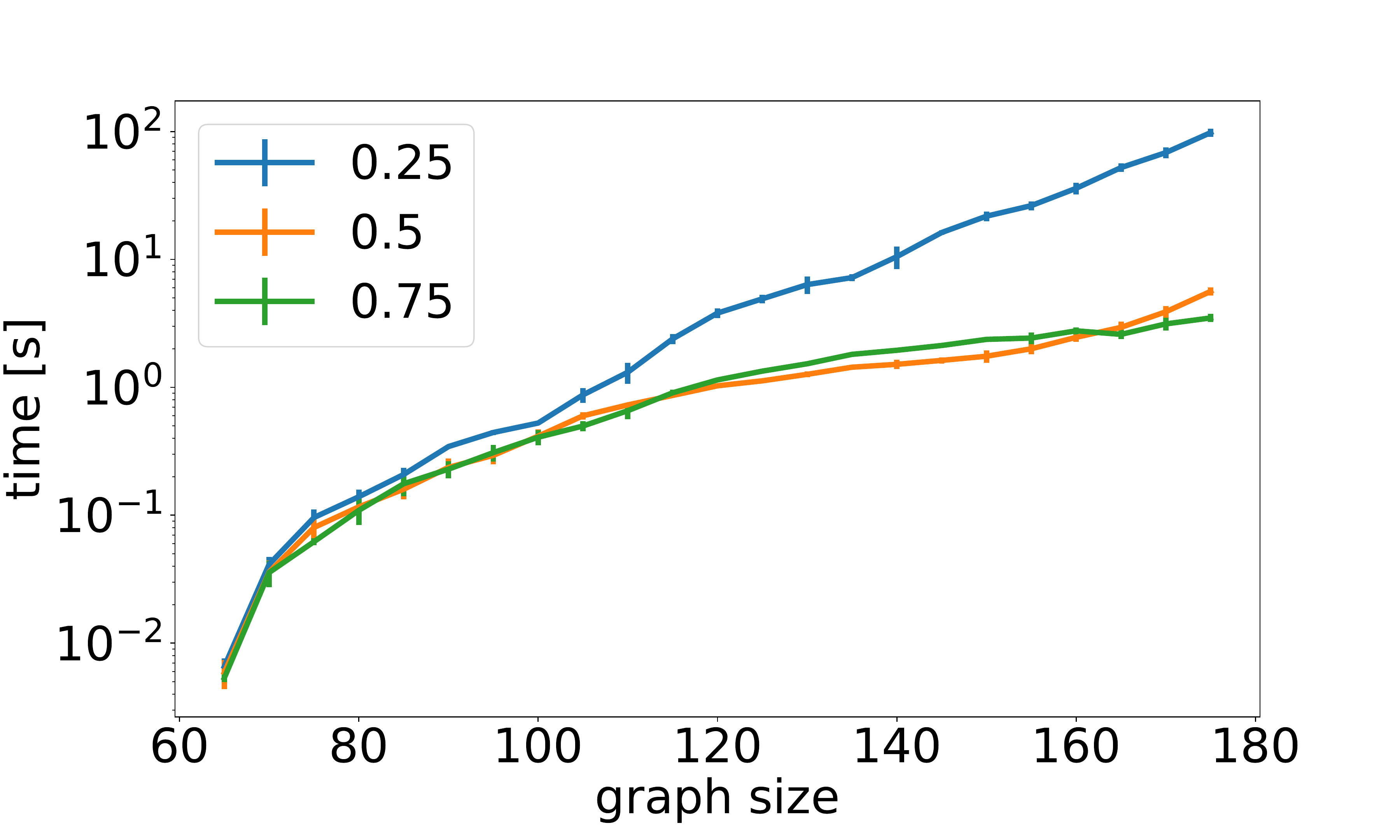}\\
    \includegraphics[width=\textwidth]{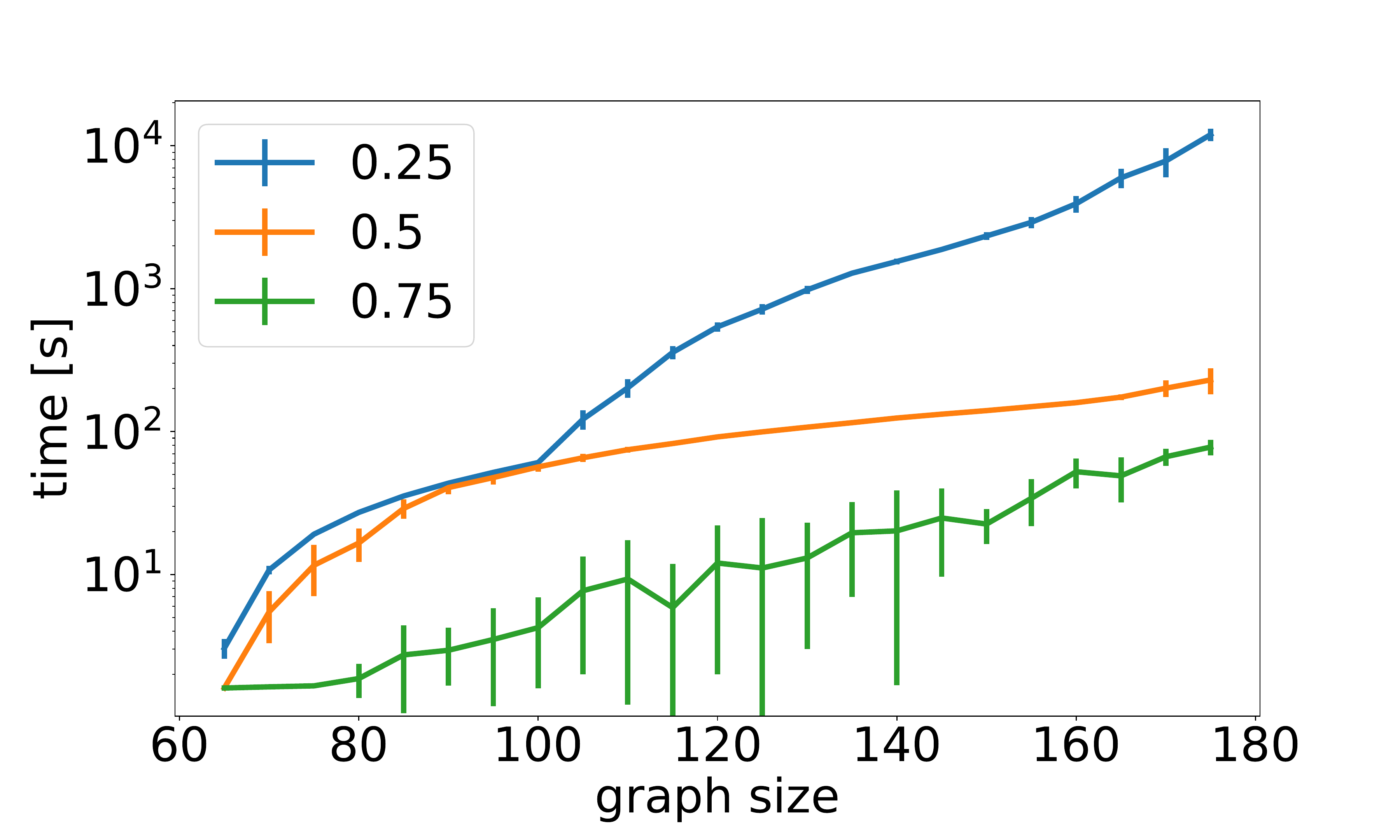}
    \caption{Performance of the DBR algorithm as a function of the graph size. Graph densities $d \in \{0.25,0.5,0.75\}$. Subgraph count (top), preprocessing time (middle) and predicted solution time (bottom) as a function of the graph density. Logarithmic scale on the y-axis. Error bars of one standard deviation.}
    \label{fig:dbr_scaling}
\end{minipage}
\end{figure}

Figure~\ref{fig:dbk_scaling} shows scaling results of DBK as a function of the graph size ranging from $60$ to $180$ vertices, and for three graph densities $d \in \{0.25,0.5,0.75\}$. We observe that the scaling is superpolynomial in the graph size, which is to be expected when solving an NP-hard problem. We also observe that as expected, denser graphs require a higher solver runtime since for dense graphs the extracted subgraphs on average contain more vertices and thus take longer to be fully decomposed.

We repeat the same scaling experiment for the MVC problem. To fully specify the algorithm of Section~\ref{sec:MVC}, we now employ the high degree vertex selection (Section~\ref{sec:experiments_vertexchoice}), the decomposition upper and chromatic lower bounds (Section~\ref{sec:experiments_bounds}), and the neighbor based vertex removal reduction strategy as determined in Section~\ref{sec:experiments_reductions}. As in \cite{acmpaper}, we call the resulting algorithm the DBR algorithm (Decomposition, Bounds, Reduction).

Analogously to the MC scaling, Figure~\ref{fig:dbr_scaling} shows scaling results for MVC. Importantly, and as expected due to the inverse relationship of the MC and MVC problems, the DBR algorithm also has a superpolynomial scaling, where higher graph densities result in a lower runtime.

\subsection{Applying our algorithms to real world graphs}
\label{sec:realworld}
\begin{table*}[t]
\centering
\begin{tabular}{|l||l|l||l|l|l|}
    \hline
    Graph name~     & No.~      & No.~      & CPU~  & No.~          & Time [s] \\
                    & vertices~ & edges~    & time~ & subgraphs~ & \\
    \hline
    bn-macaque-rhesus-interareal- & 93 & 2700 & 0.0701 & 1 & 1.670 \\
    cortical-network-2 &&&&&\\
    ENZYMES-g8 & 88 & 133 & 0.016 & 1 & 1.616 \\
    ENZYMES-g123 & 90 & 127 & 0.0167 & 1 & 1.617 \\
    rt-retweet & 96 & 117 & 0.0199 & 1 & 1.619 \\
    polbooks & 105 & 441 & 0.0329 & 1 & 1.633 \\
    ia-enron-only & 143 & 623 & 0.0756 & 1 & 1.676 \\
    ia-infect-hyper & 113 & 2196 & 0.0846 & 1 & 1.685 \\
    johnson16-2-4 & 120 & 5460 & 2.611 & 531 & 852.211 \\
    \hline
\end{tabular}
\caption{Predicted solution time in seconds for real world graphs based on a single run using DBK. \label{table:real_graphs_dbk}}
\end{table*}
To demonstrate the applicability of our proposed algorithms, we apply them to find cliques or vertex covers in real world graphs \citep{nr, rossi2014pmc-www, rossi2012fastclique, cohen2005enron, bigbrain, infect, rossi2014coloring, dimacs}. For MC, Table~\ref{table:real_graphs_dbk} shows results for the DBK algorithm, demonstrating that graphs with hundreds of vertices and thousands of edges can be solved in a few seconds. We also observe that the bounding and reduction techniques result in a strong pruning of the generated subproblems, since the number of created subproblems is typically very low.

\begin{table*}[t]
\centering
\begin{tabular}{|l||l|l||l|l|l|}
    \hline
    Graph name~     & No.~      & No.~      & CPU~  & No.~          & Time [s] \\
                    & vertices~ & edges~    & time~ & subgraphs~ & \\
    \hline
    bn-macaque-rhesus-interareal- & 93 & 2700 & 0.104 & 1 & 1.704 \\
    cortical-network-2 &&&&&\\
    ENZYMES-g8 & 88 & 133 & 0.0233 & 2 & 3.223 \\
    ENZYMES-g123 & 90 & 127 & 0.0205 & 1 & 1.620 \\
    rt-retweet & 96 & 117 & 0.0161 & 1 & 1.602 \\
    polbooks & 105 & 441 & 0.060 & 1 & 1.660 \\
    ia-enron-only & 143 & 623 & 0.169 & 6 & 9.769 \\
    ia-infect-hyper & 113 & 2196 & 0.350 & 22 & 35.550 \\
    johnson16-2-4 & 120 & 5460 & 0.536 & 2 & 3.736 \\
    \hline
\end{tabular}
\caption{Predicted solution time in seconds for real world graphs based on a single run using DBR. \label{table:real_graphs_dbr}}
\end{table*}
For MVC, an assessment of the DBR algorithm in Table~\ref{table:real_graphs_dbr} confirms these results.

\subsection{Performance on future D-Wave architectures}
\label{sec:experiments_future}
When using our algorithms in connection with the D-Wave annealer, DBR or DBK will be run until the size of a subproblem created during the decomposition reaches at most $46$ vertices for D-Wave 2X ($64$ for D-Wave 2000Q, and $180$ for D-Wave Advantage), since QUBOs of this size are guaranteed to be embeddable on the qubit architectures. It is interesting to investigate how the scaling behavior of our algorithms depends on this cutoff of the decomposition. For this we apply our DBR and DBK algorithms to random graph instances of fixed density, and decompose those graphs until a limit is reached that depends on the D-Wave architectures we investigate. This allows us to report numbers of generated subgraphs (subgraph count) for each architecture and preprocessing times. Assuming a fixed time of $1.6$ seconds for $10000$ anneals as for D-Wave 2X and 2000Q, we can also report predicted solution times for D-Wave Advantage.

\begin{figure}
\begin{minipage}{0.49\textwidth}
    \centering
    \includegraphics[width=\textwidth]{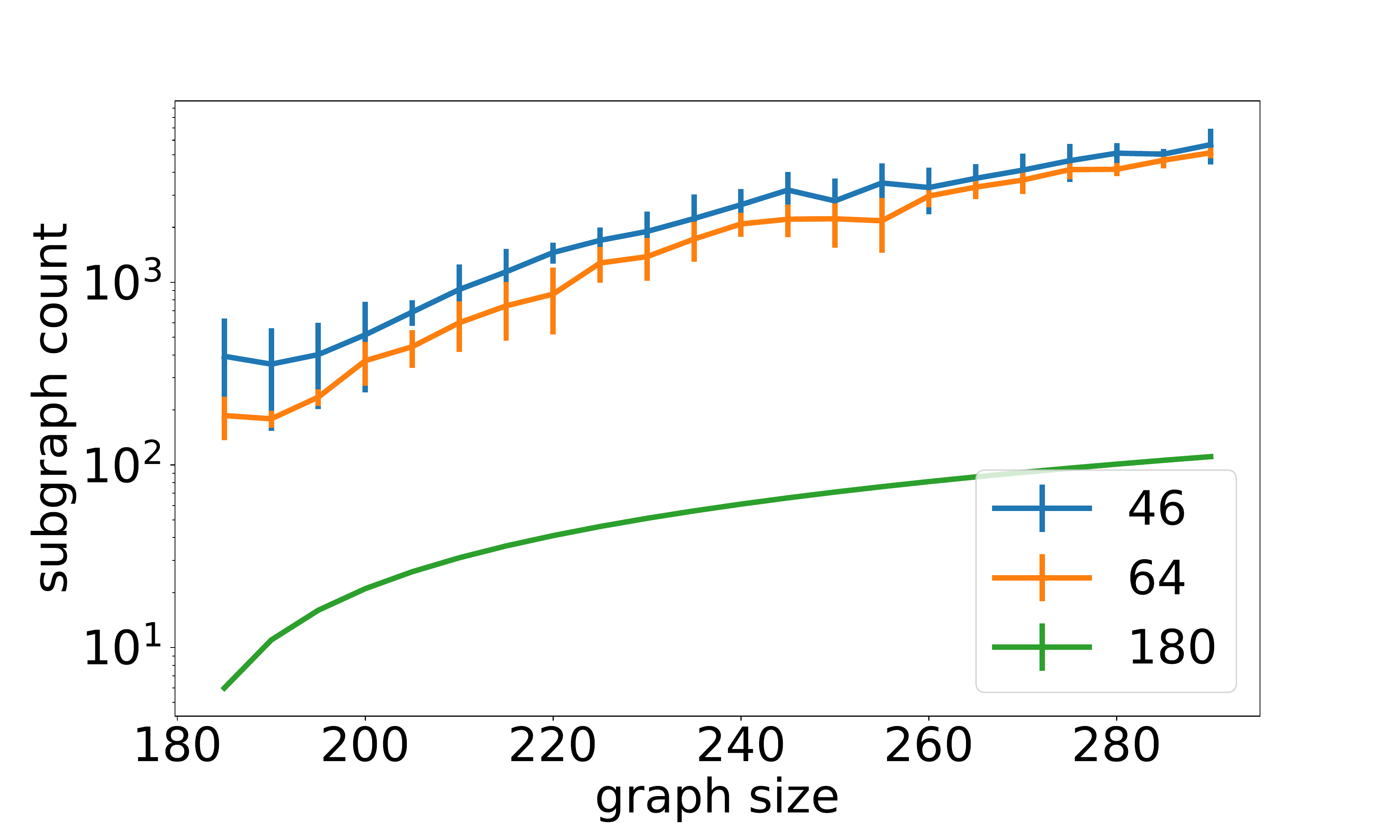}\\
    \includegraphics[width=\textwidth]{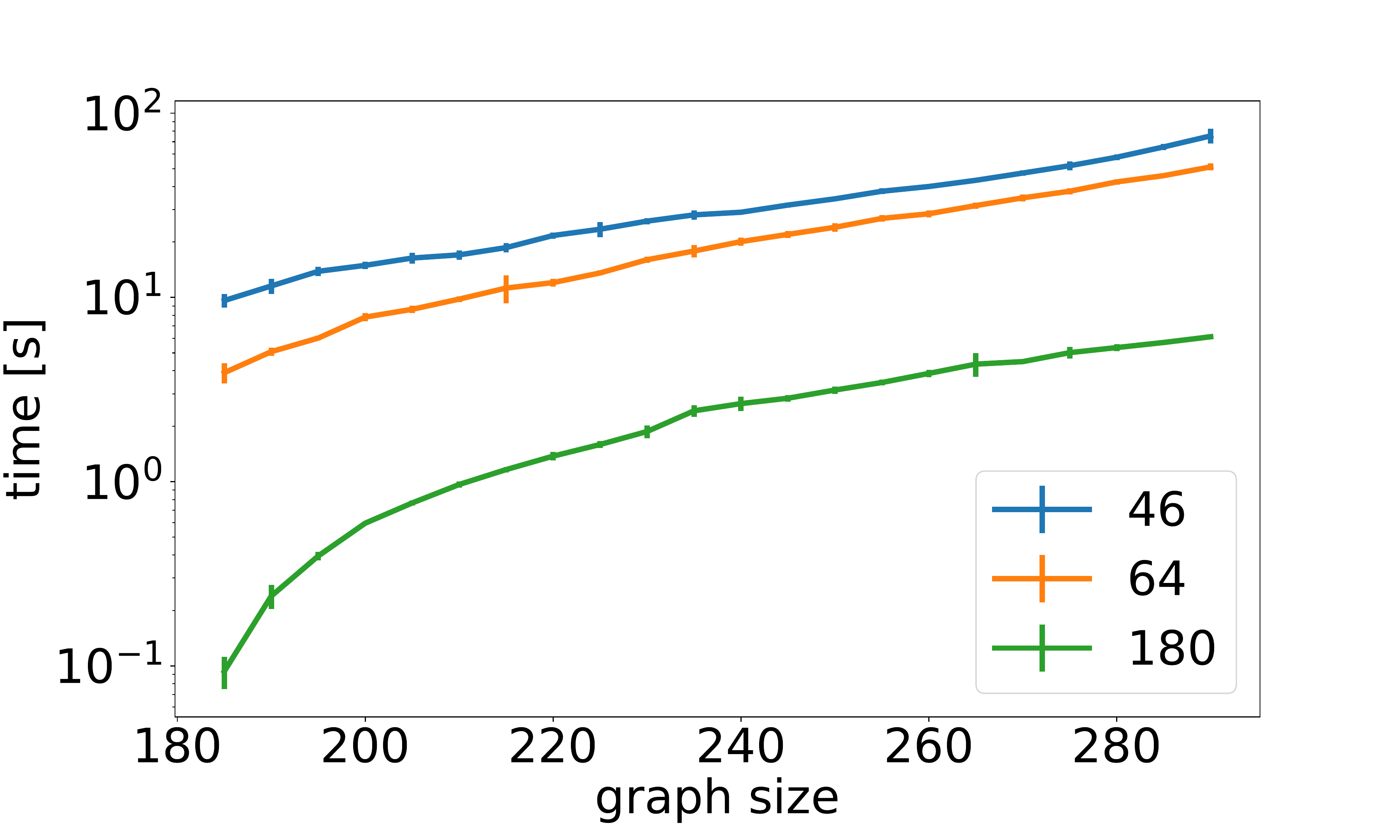}\\
    \includegraphics[width=\textwidth]{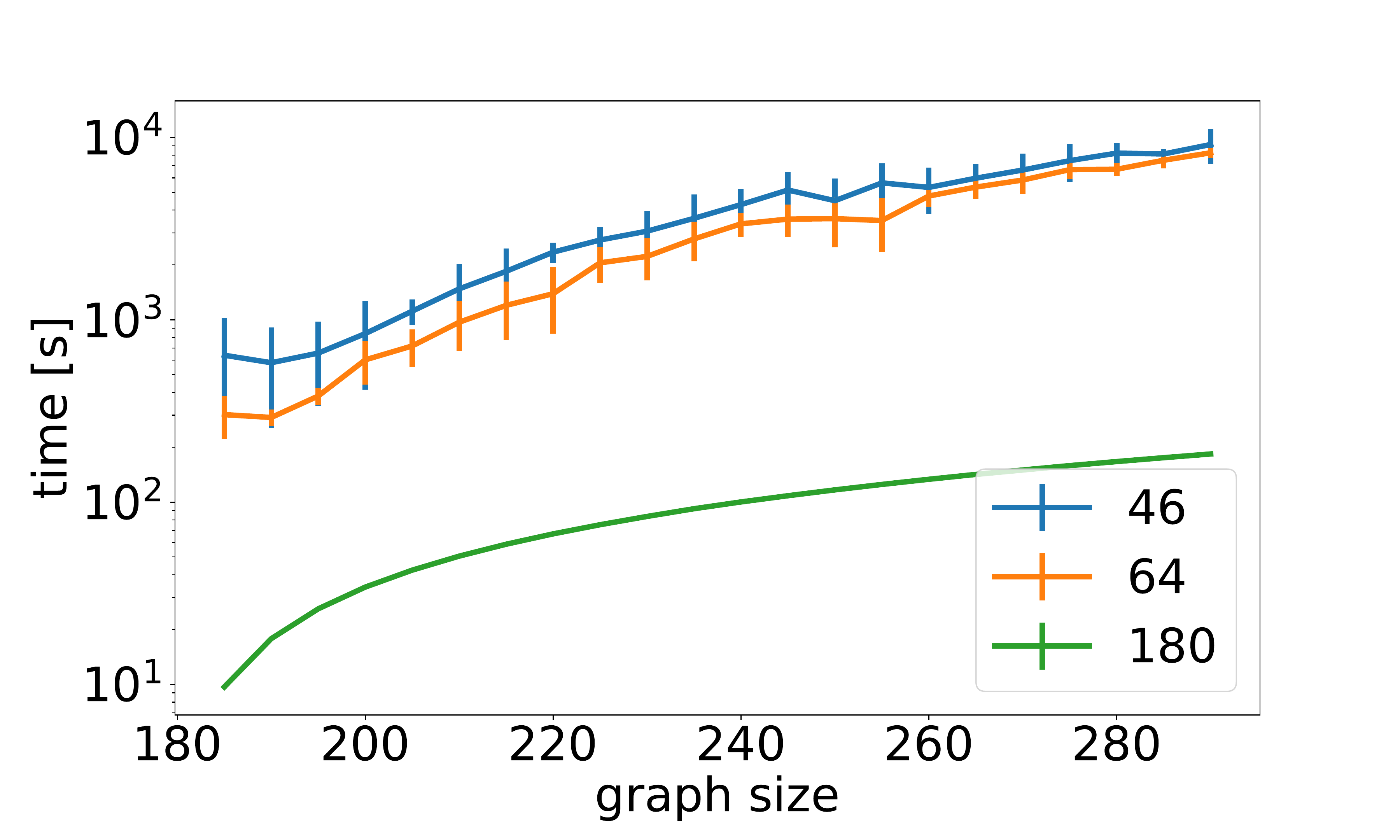}
    \caption{Performance prediction of the DBK algorithm on future D-Wave architectures as a function of the graph size. Recursion cutoff at subgraph sizes of $46$, $64$, and $180$ vertices. Subgraph count (top), preprocessing time (middle) and predicted solution time (bottom) as a function of the graph size. Logarithmic scale on the y-axis. Error bars of one standard deviation. Graph density $0.5$.}
    \label{fig:dbk_future}
\end{minipage}\hfill
\begin{minipage}{0.49\textwidth}
    \centering
    \includegraphics[width=\textwidth]{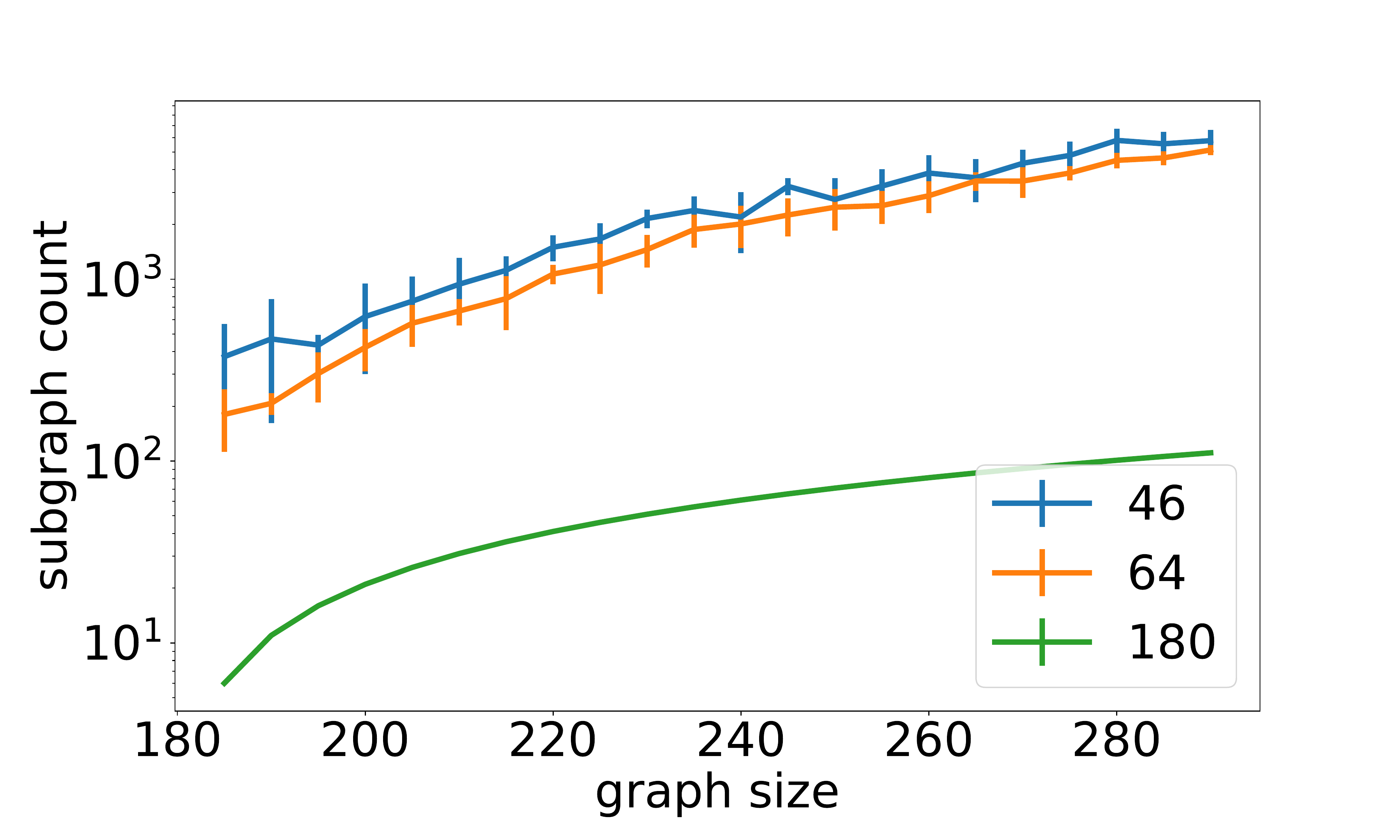}\\
    \includegraphics[width=\textwidth]{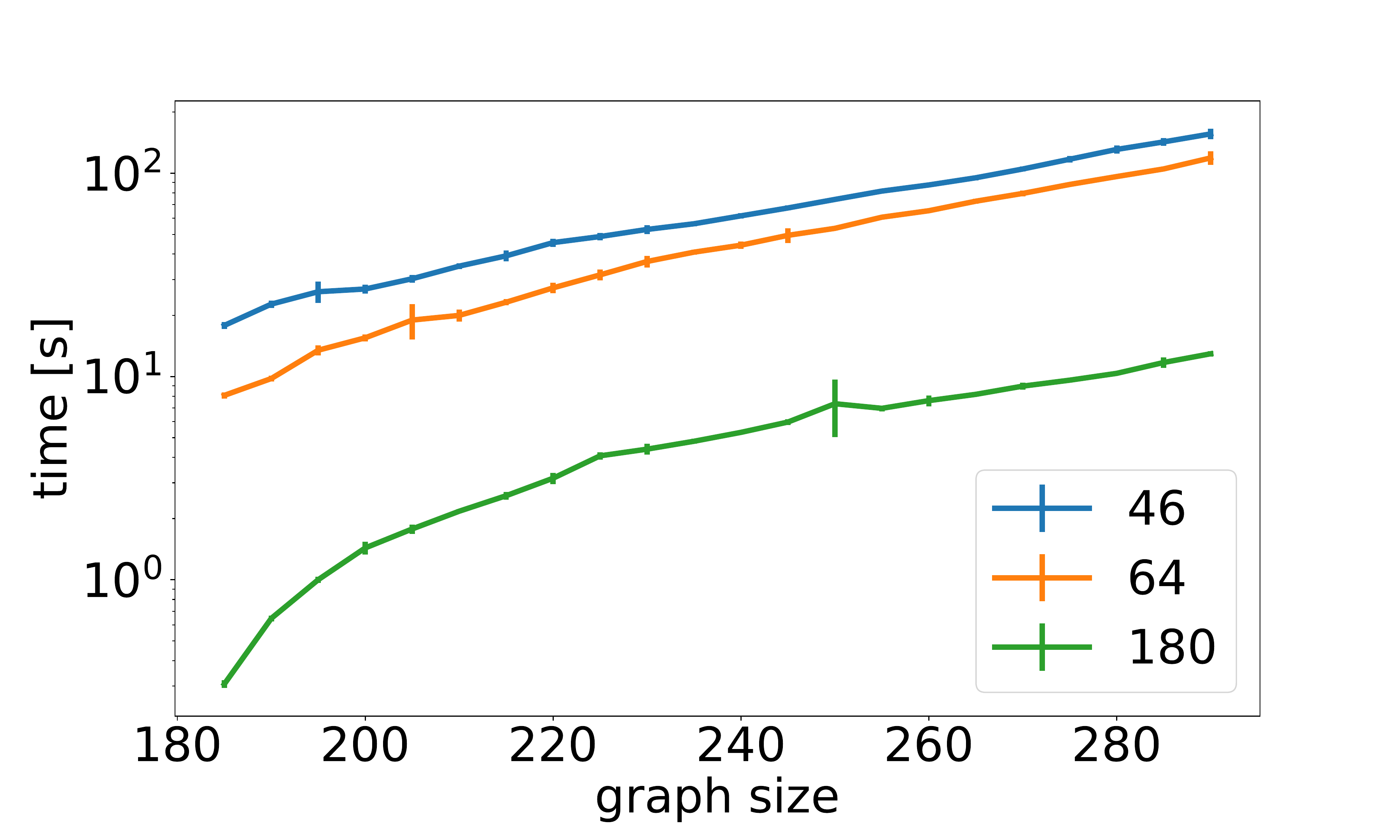}\\
    \includegraphics[width=\textwidth]{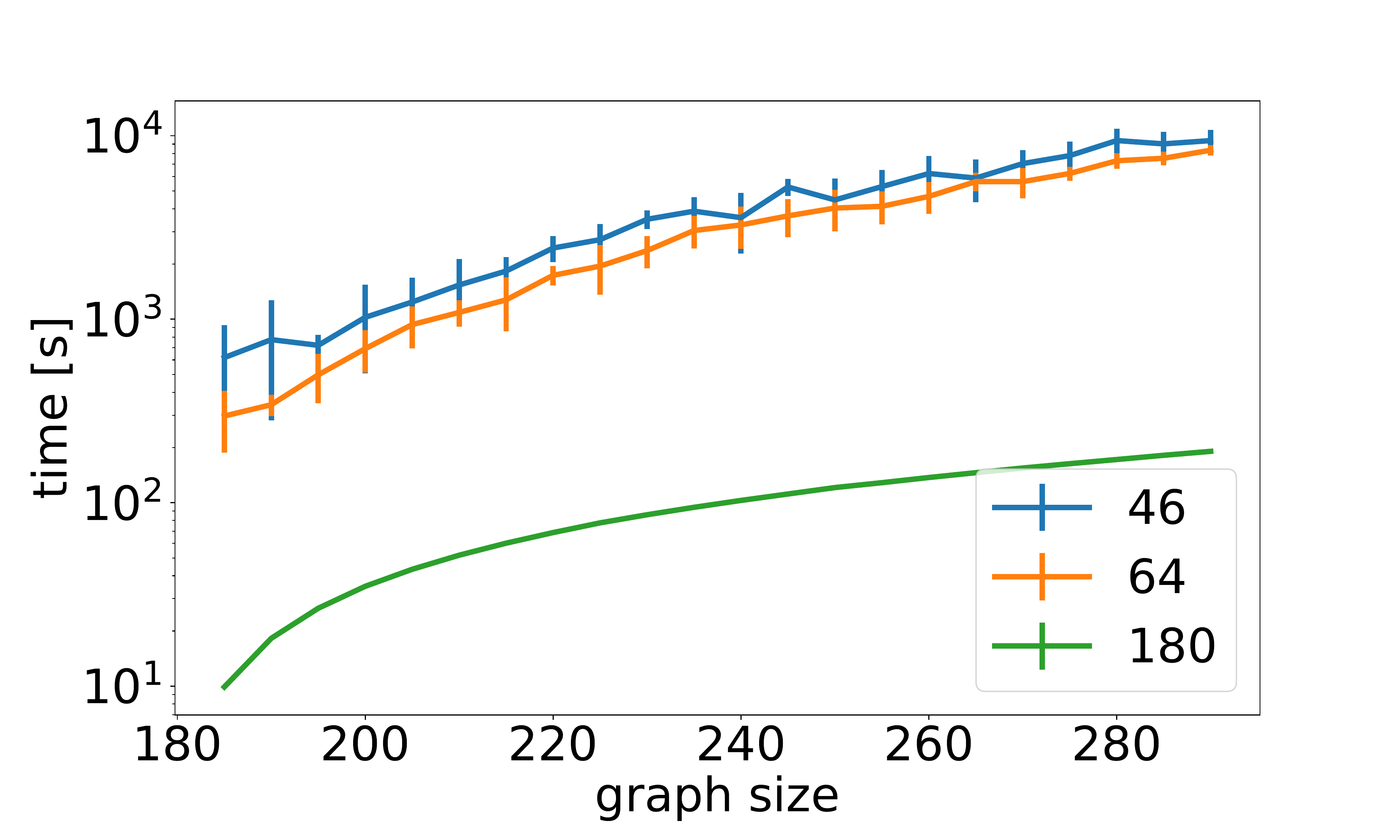}
    \caption{Performance prediction of the DBR algorithm on future D-Wave architectures as a function of the graph size. Recursion cutoff at subgraph sizes of $46$, $64$, and $180$ vertices. Subgraph count (top), preprocessing time (middle) and predicted solution time (bottom) as a function of the graph size. Logarithmic scale on the y-axis. Error bars of one standard deviation. Graph density $0.5$.}
    \label{fig:dbr_future}
\end{minipage}
\end{figure}

For a fixed graph density of $0.5$, Figure~\ref{fig:dbk_future} shows runtime predictions for DBR. As expected, a higher cutoff leads to a faster runtime. Notably, we observe almost no difference between D-Wave 2X and 2000Q, but a pronounced speedup on D-Wave Advantage. It seems as if the slope for the $180$ vertex cutoff on D-Wave Advantage slightly decreases, but this remains for further investigation.

Prediction results for the DBR algorithm (Figure~\ref{fig:dbr_future}) are qualitatively similar.

\section{Discussion}
\label{sec:discussion}
This article proposed a novel decomposition framework for NP-hard graph problems characterized by finding an optimal set of vertices. The framework recursively splits a given instance of a NP-hard graph problem into smaller subproblems until, at some recursion level, the generated subproblems can be solved with any method of choice. This includes but is not limited to a quantum annealer such as the ones of D-Wave, Inc. The algorithm is exact, meaning that the optimal solution of the original problem is guaranteed under the assumption that all subproblems are solved exactly.

We concretize our framework for two important NP-hard graph problems, the Maxmimum Clique (MC) and the Minimum Vertex Cover (MVC) problems. In both cases, we arrive at a decomposition method capable of splitting arbitrarily large problem instances into subproblems solvable on D-Wave.

To speed up the computations, our generic algorithm allows for the specification of bounds and reduction techniques which help to reduce the computational workload. We investigate several such techniques in detail in the experimental analysis section, and use our results to fully specify the DBK (for MC) and DBR algorithms (for MVC). We summarize our findings as follows:
\begin{enumerate}
    \item Our results nicely confirm the inverse relationship of the MC and MVC problems in that empirically, the best lower (upper) bounds are also the best upper (lower) bounds of the other problem. Moreover, the scaling behavior of our algorithms as a function of the graph density is nicely inverted.
    \item Both algorithms show a reasonable scaling behavior and exactly solve graphs with about $300$ vertices in less than one hour. We show that an application of our methods to real world graphs is feasible.
    \item A performance prediction on future D-Wave architectures shows that our algorithms will behave very favorably on future annealer generations, in the sense that a higher qubit connectivity on the D-Wave chip will result in a considerable runtime reduction for our methods.
\end{enumerate}

With this article we solely aim to provide a method that can be used to help quantum annealers solve problems which are too large to be implemented onto their hardware. Current quantum devices are in their infancy, and they are neither large enough nor accurate enough to compete with classical computers at solving (general) optimization problems. Therefore, we do not aim to compete with current state-of-the-art classical solvers. However, no matter how large and how accurate quantum computers become in the future, there will always exist problems too large for their hardware, and decomposition algorithms like the proposed ones will be needed.

Future work includes the investigation of further techniques to bound, reduce and prune the subproblems created during the decomposition. Moreover, more NP-hard problems could be investigated with our framework, and an improved implementation of our DBK and DBR algorithms would be beneficial.

\section*{Acknowledgments}
\label{sec:Acknowledgments}
Research presented in this article was supported by the Laboratory Directed Research and Development program of Los Alamos National Laboratory under project numbers 20180267ER and 20190065DR.


\end{document}